\documentclass[%
aip,
jcp,
reprint,
graphicx,
amsmath,amssymb,amsthm
longbibliography, 
nofootinbib 
]{revtex4-1}

\usepackage[%
bookmarks,%
bookmarksopen=false,
pdfauthor={Autor},%
pdftitle={Titel},%
colorlinks=true,
linkcolor=blue,
citecolor=red,%
urlcolor=cyan,%
]{hyperref}

\usepackage{array}
\usepackage{refcount}
\usepackage{dcolumn}
\usepackage{bm}
\usepackage{braket}
\usepackage[T1]{fontenc}
\usepackage[utf8]{inputenc}
\usepackage[english]{babel}
\usepackage{comment}
\usepackage{amssymb} 
\usepackage{caption}
\usepackage{subcaption}

\usepackage{float}
\usepackage{booktabs}
\usepackage{multirow, makecell}

\usepackage{chemfig}
\usepackage{chemexec}
\usepackage{tikz}
\usetikzlibrary{positioning, calc}
\usepackage{pgfplots}
\pgfplotsset{compat=1.17}
\usepackage{derivative}
\usepackage[abs]{overpic}
\usepackage[T1]{fontenc}
\usepackage{epsfig}
\usepackage{dsfont}
\usepackage{subcaption}
\usepackage{ulem}
\usepackage{cancel}
\usepackage{xcolor}

\usepackage{moreverb,rotating,graphics}
\usepackage{epsfig}
\usepackage{extarrows}
\usepackage{dsfont}
\usepackage{mathtools}
\usepackage{threeparttable}
\usepackage{pifont}

\allowdisplaybreaks[4]

\newcommand{\tabincell}[2]{\begin{tabular}{@{}#1@{}}#2\end{tabular}}

\newcolumntype{C}[1]{>{\centering\arraybackslash}p{#1}}
\newcolumntype{L}[1]{>{\raggedright\arraybackslash}p{#1}}
\newcolumntype{R}[1]{>{\raggedleft\arraybackslash}p{#1}}

\def\R{{\mathbb R}}
\def\C{{\mathbb C}}

\def\1{{\mathds{1}}}

\begin{document}

	
	
	\title{Critical point search and linear response theory for computing electronic excitation energies of molecular systems. $\mbox{Part II: CASSCF}$}
	\author{Laura Grazioli}
	\email{laura.grazioli@enpc.fr}
	\affiliation{CERMICS, CNRS, Ecole des Ponts, Institut Polytechnique de Paris and Inria, 6-8 avenue Blaise Pascal, Cit\'e Descartes, 77455 Marne-la-Vall\'ee, France}
	\author{Yukuan Hu}%
	\email{yukuan.hu@enpc.fr}
	\affiliation{CERMICS, CNRS, Ecole des Ponts, Institut Polytechnique de Paris, 6-8 avenue Blaise Pascal, Cit\'e Descartes, 77455 Marne-la-Vall\'ee, France}
	\author{Tommaso Nottoli}%
	\email{tommaso.nottoli@dcci.unipi.it}
	\affiliation{Dipartimento di Chimica e Chimica Industriale, Università di Pisa, I-56124 Pisa, Italy}
	\author{Filippo Lipparini}%
	\email{filippo.lipparini@unipi.it}
	\affiliation{Dipartimento di Chimica e Chimica Industriale, Università di Pisa, I-56124 Pisa, Italy}
	\author{Eric Canc\`es}
	\email{eric.cances@enpc.fr}
	\affiliation{CERMICS, CNRS, Ecole des Ponts, Institut Polytechnique de Paris and Inria, 6-8 avenue Blaise Pascal, Cit\'e Descartes, 77455 Marne-la-Vall\'ee, France}
	
	
	\date{\today}
	
	\begin{abstract}
		The computation of excited states within the Complete Active Space Self-Consistent Field (CASSCF) framework remains a significant challenge in quantum chemistry, both theoretically and algorithmically. In this work, we extend the K\"ahler manifold formalism introduced in Part I of this series to the CASSCF theory, and draw a geometrical connection from the time-dependent CASSCF equations to state-specific and linear response methodologies for excited states. This is achieved by first investigating the underlying CASSCF manifold and identifying its K\"ahler structure, which is complicated by the nontrivial coupling of CI and orbital degrees of freedom. Building on these theoretical findings, we derive the CASSCF linear response equations in a straightforward manner, and develop a robust state-specific method that relies solely on first-order derivatives of the CASSCF energy functional. Numerical results on representative molecular systems---water, formaldehyde, and ethylene---demonstrate the effectiveness of the proposed state-specific method, while revealing the difficulty of reliable identification of excited states due to nonlinearity induced by the CASSCF theory. 
	\end{abstract}
	
	
	\maketitle 
	
	
	\section{Introduction}
	The Complete Active Space Self-Consistent Field (CASSCF) is a powerful method to qualitatively describe strongly correlated molecular systems.\cite{roos1980complete,roos1980complete2,roosmulticonfigurational,Werner1987,Shepard1987,Roos1987} It offers access to both ground and excited states, and is therefore a popular choice for investigating photophysical and photochemical processes. However, applying CASSCF suffers from many limitations. The two most apparent problems of CASSCF are its computational cost, given the exponential scaling of the underlying CASCI problem, and the arbitrariness introduced by the choice of the active space. In other words, CASSCF calculations are sometimes more an art than a science, even though a lot of progress has been made to formulate robust algorithms\cite{Werner1980,Werner1985,jensen1984direct,jensen1986direct,Sun17,Kreplin2019} and automated active space selection strategies.\cite{Pulay1988,toth2020comparison,Stein2016,Sayfutyarova2017} 
	
	There are, however, several other aspects that make CASSCF a method that is far from being straightforward or black-box. From a numerical point of view, the optimization of the CASSCF wavefunction is a challenging problem even for ground-state calculations. This is mainly due to the fact that CASSCF requires one to optimize at the same time both the molecular orbitals and the CI coefficients. These are two very different tasks, the former being a highly nonlinear optimization problem of moderate size, and the latter being a very high-dimensional linear eigenvalue problem. What really makes CASSCF optimization hard is that the two tasks are not independent, and the subtle interplay between the two sets of degrees of freedom can lead to poor convergence or instabilities.\cite{Werner1980,Werner1985}
	
	Computing excited states is an even more challenging task. As CASSCF is a nonlinear method, the definition of excited state itself is not straightforward, i.e., there is no corresponding linear and self-adjoint Hamiltonian operator whose eigenfunctions are the excited states. 
	
	The most popular approach, state-averaged CASSCF,\cite{werner1981quadratically} optimizes the orbitals for an average of CASCI electronic states that are obtained by seeking the few lowest eigenvalues of the CASCI Hamiltonian. State-averaged CASSCF has the advantage of being relatively straightforward to implement starting from an available ground-state CASSCF implementation, but limits the flexibility of the method by relegating the treatment of the electronic excitation to the active space, and produces orbitals that are not optimal for any particular state.\cite{helmich2019benchmarks} Nevertheless, it represents a good compromise between cost-effectiveness, simplicity, and accuracy.
	
	Two more approaches can be used to compute excited states with CASSCF. State-specific calculations\cite{hanscam2022applying,helmich2022trust,jensen1986direct,jensen1984direct,tran2020improving,tran2019tracking} can be performed, where one optimizes a CASSCF wavefunction that corresponds to a saddle point of the CASSCF energy as a function of the variational parameters. Although this sounds like a straightforward definition, it comes with some problems. First, it is not obvious that all the saddle points of the CASSCF energy correspond to physical states, due to the nonlinearity of the model. Second, saddle point optimizations are, in general, more difficult than minimizations. \\
	Another popular approach is given by linear response theory,\cite{dalgaard1980time,grazioli2026critical,helmich2019casscf,jorgensen1988linear,olsen1985linear,yeager1979multiconfigurational} where excitation energies are obtained from poles of the molecular linear response function. This is a rigorous and well-defined approach, but the so-obtained excitation energies may strongly depend on the CASSCF ground state around which the dynamics is linearized. Furthermore, it can be computationally demanding, and the numerical methods for solving linear response equations can encounter convergence difficulties, especially if many states are sought.\cite{helmich2019casscf,alessandro2023linear,nottoli2025efficient}
	
	Recently, real-time time-dependent CASSCF has been proposed as a method to investigate electronic dynamics. Electronic excitation energies, and even whole spectra, can also be obtained from real-time simulations by analyzing appropriate correlation functions.\cite{li2024implementation,sato2013time,sato2016time}
	
	In summary, not only is CASSCF a complex method \textit{per se}, but its extension to excited states introduces significant complications, even speaking of the strategy used to define them. 
	In this paper, we aim to shed some light on the method by exploring and clarifying the geometrical structure of CASSCF, and by using such a structure to connect, in a clear, unambiguous way, the CASSCF dynamics (i.e., time-dependent formulation) to the solution of the CASSCF equations for ground and excited states in a state-specific context, and to CASSCF linear response theory. 
	
	We achieve these by recognizing and exploring the K\"ahler structure of the CASSCF manifold, which 
	provides a clear path to deriving time-dependent CASSCF equations. The K\"ahler formalism was introduced by Erich K\"ahler in the 1930s to unify holomorphic geometry, Riemannian geometry and Hamiltonian dynamics.\cite{kahler1933uber} It also plays a key role in theoretical physics, for example in Calabi-Yau manifolds,\cite{yau2008survey} supersymmetry and string theory.\cite{greene1997string} Building on the K\"ahler structure, we provide a new, more geometrical way of deriving CASSCF linear response equations, corresponding to the linearization of the 
	CASSCF dynamics around the CASSCF ground state. We also show that the stationary points of the CASSCF dynamics correspond to (state-specific) ground and excited states, and develop a constrained saddle-search method, which robustly locates the saddle points on the CASSCF energy landscape with any given Morse index and requires only first-order derivatives of the energy functional. 
	
	Understanding the geometrical structure of CASSCF is not straightforward. If one decouples the CI problem from the orbital optimization, one obtains two problems for which the geometrical structure is already well defined. The CI problem corresponds in fact to the minimization of a quadratic function on the unit sphere, which leads to the CASCI linear eigenvalue problem. The orbital optimization problem, on the other hand, corresponds to a highly nonlinear optimization problem on a flag manifold, which is more complex, but already well understood.\cite{ye2022optimization,vidal2024geometric} 
	%
	%
	The real challenge is to understand the full picture, where both the orbitals and CI coefficients are variable, as the resulting manifold is not just the simple union of the two aforementioned manifolds, but a more complex entity that allows for the subtle interplay of the two sets of degrees of freedom, which is what makes CASSCF such a complicated method. We note that in Part I of the series of contributions, some of the authors adopted the K\"ahler formalism for the Full CI, Hartree-Fock methods (HF), and Density Functional Theory (DFT), where the simpler complex Grassmann manifolds are involved.\cite{grazioli2026critical}

	We believe that the theoretical developments in this paper will open new avenues to formulating new numerical strategies for CASSCF excited-state calculations in the state-specific framework, as well as a better understanding of the problem of whether a critical point actually corresponds to a physically-relevant excited state, or is a spurious state which stems from the nonlinearity of the CASSCF approximation.  
	The cases of other advanced variational approximations such as Density Matrix Renormalization Group (DMRG)\cite{white1992density,white1993density} and bivariational approximations such as Coupled Cluster (CC)\cite{bartlett2007coupled,vcivzek1991origins,cizek1980coupled} will be dealt with in Parts III and IV.
	
	\par This paper is organized as follows. In Section \ref{sec:CASSCF manifold Kahler}, we elaborate on the CASSCF manifold and identify its K\"ahler structure, where the coupling orbitals and CI degrees of freedom are treated. Based on this, in Section \ref{sec:TD CASSCF}, we write out the closed forms of Riemannian gradient and Hessian of the CASSCF energy functional, and give the CASSCF dynamics (i.e., time-dependent CASSCF equations). In Section \ref{sec:CASSCF LR}, the CASSCF linear response theory is investigated by linearizing the CASSCF dynamics around the CASSCF ground state. In Section \ref{sec:CASSCF SS}, we recognize the CASSCF ground and excited states as the stationary points of the CASSCF dynamics and propose a constrained gentlest ascent method for locating CASSCF excited states. In Section \ref{sec:numerics}, we test the proposed method in finding state-specific excited states of water, formaldehyde, and ethylene molecules, implemented with an interface to the CFOUR program.\cite{cfour} 
	The state-specific results are compared with those from CASSCF state-averaged and linear response calculations.

	\section{The set of CASSCF states and its K\"ahler structure}\label{sec:CASSCF manifold Kahler}
	
	\subsection{CASSCF wavefunctions and CASSCF state space}
	
	Let $\{\chi_\mu\}_{\mu=1}^{N_{\rm b}}$ be an orthonormal set of atomic orbitals, and let
	\[
	\phi_p^C := \sum_{\mu=1}^{N_{\rm b}} C_{\mu p}\chi_\mu,\quad p=1, \ldots, N_{\rm b}
	\]
	be an orthonormal set of molecular orbitals (MOs) defined as linear combinations of atomic orbitals with coefficients $C$, where $C\in U(N_{\rm b})$ and $U(N_{\rm b})$ denotes the unitary group of degree $N_{\rm b}$. 
	In CASSCF, we introduce a partition of the molecular orbitals into three families. The $N_{\rm I}\in\mathbb{N}$ internal orbitals are always (doubly, in the spin case) occupied in the CASSCF wavefunction, and are denoted with subscripts $i,j,\ldots$. The $N_{\rm E}\in\mathbb{N}$ external orbitals are always empty in the CASSCF wavefunction, and are denoted with subscripts $a, b, \ldots$. Finally, the $N_{\rm A}\in\mathbb{N}$ active MOs have variable occupation numbers and are denoted with subscripts $u,v,\ldots$. The CASSCF wavefunction is generated by distributing the $n_{\rm a}\in\mathbb{N}$ active electrons into the $N_{\rm A}$ active MOs in a Full CI fashion, i.e., by generating all possible configurations. 
	
	The CASSCF wavefunction therefore reads
	\begin{equation*}
		\begin{split}
			|\Psi_{c,C}\rangle := &~\sum_{1\le u_1^-<u_2^-<\cdots<u_{n_{\rm a}}^-\le N_{\rm A}}c_{u_1^-,u_2^-,\ldots, u_{n_{\rm a}}^-} \\
			&\qquad\qquad\qquad\qquad\widehat a^{\dagger,C}_{u_1}\widehat a^{\dagger,C}_{u_2}\ldots \widehat a^{\dagger,C}_{u_{n_{\rm a}}} | - \rangle^C,
		\end{split}
	\end{equation*}
	Here the notation $u^-:=u-N_{\rm I}$ denotes the shifted index for the $u$-th MO, which is in fact the $u^-$-th active MO ($u=N_{\rm I}+1,\ldots,N_{\rm I}+N_{\rm A}$). The operators $\{\widehat a_p^{\dagger,C}\}_{p=1}^{N_{\rm b}}$ are the creation operations in the formalism of second quantization relative to the basis $\{\phi^C_p\}_{p=1}^{N_{\rm b}}$. The parameter $c_{u_1^-,u_2^-,\ldots,u_{n_a}^-}\in\mathbb C$ is the CI coefficient associated with the Slater determinant 
	containing the active MOs indexed by $u_1^-,u_2^-,\ldots,u_{n_a}^-$. Note that the creation operators in the definition of the Slater determinants that constitute the CASSCF wavefunction are restricted to act on active MO indices only, and that the vacuum state $|-\rangle^C$ is defined as the Slater determinant consisting of $N_{\rm I}$ internal orbitals.
	
	For the sake of compactness, we regroup all the $u_1^-,u_2^-,\ldots, u_{n_{\rm a}}^-$ indices into the multi-index $A$ and write
	\[
	|\Psi_{c,C}\rangle = \sum_{A=1}^{N_{\rm det}} c_A |A^C\rangle,
	\]
	where $N_{\rm det}\in\mathbb{N}$ is the number of Slater determinants $|A^C\rangle$ defined from the MO basis in the CAS space; more precisely, $N_{\rm det}=\binom{N_{\rm A}}{n_{\rm a}}$. The CI coefficients $c_A$'s satisfy
	\[
	\sum_{A=1}^{N_{\rm det}} |c_A|^2 = 1,
	\]
	or equivalently, $c\in\mathbb{S}_{\C}^{N_{\det}-1}$, where $\mathbb{S}_{\C}^{N_{\det}-1}$ denotes the complex unit sphere in $\C^{N_{\det}}$. 
	
	The CI coefficients $c$ and MO coefficients $C$ fully describe a CASSCF wavefunction $\ket{\Psi_{c,C}}$. Clearly, $(c,C)\in\mathbb{S}_{\mathbb{C}}^{N_{\rm det}-1}\times U(N_{\rm b})$. In addition, we say that two elements $(c,C)$ and $(c',C')$ are equivalent, which we denote by $(c,C) \sim (c',C')$, if they generate the same wavefunction up to a global phase factor. This is achieved when the following relations hold
	\begin{equation*}
		C'=C\begin{pmatrix}
			U_{\rm I} & 0 & 0\\
			0 & U_{\rm A} & 0\\
			0 & 0 & U_{\rm E}
		\end{pmatrix},~\begin{array}{l}
			c' =  \lambda  c \ltimes U_{\rm A}, \quad U_{\bullet} \in  U(N_\bullet),\\[0.2cm]
			\bullet \in \{{\rm I},{\rm A},{\rm E}\}, \quad \lambda \in  U(1),
		\end{array}
	\end{equation*}
	where the notation $c'=\lambda c \ltimes U_{\rm A}$ means for any $1\le u_1^-<\cdots<u_{n_{\rm a}}^-\le N_{\rm A}$,
	\begin{align}
		c'_{u_1^-,\cdots, u_{n_{\rm a}}^-} =&~\lambda \sum_{1\le v_1^-<\cdots<v_{n_{\rm a}}^-\le N_{\rm A}}\sum_{\sigma\in\mathcal P_{n_{\rm a}}}{\rm sgn}(\sigma)
		c_{v_1^-,\ldots, v_{n_{\rm a}}^-}\nonumber\\
		&[U_{\rm a}]_{v_1^-,u_{\sigma(1)}^-}^* \cdots [U_{\rm a}]_{v_{n_{\rm a}}^-,u_{\sigma(n_{\rm a})}^-}^*.
		\label{eqn:quotient structure transformation on c}
	\end{align}
	The notation $\mathcal P_{n_{\rm a}}$ refers to the permutation group related to $\{1,\ldots,n_{\rm a}\}$ and ${\rm sgn}(\sigma)$ gives the parity of the permutation $\sigma$. The CASSCF state space $\mathcal X$ can be defined as the set of equivalence classes induced by this relation:
	$$
	\mathcal X := \left\{ (c,C) \in \mathbb S_\C^{N_{\rm det}-1} \times U(N_{\rm b})\right\}/\sim \; .
	$$
	
	\subsection{CASSCF manifold and K\"ahler structure}
	
	Let $\gamma^{\rm act}\in\C_{\rm herm}^{N_{\rm A}\times N_{\rm A}}$ be the CASSCF active one-body reduced density matrix (1-RDM) in the MO basis, defined as
	\[
	\left[\gamma^{\rm act}_{c,C}\right]_{u^-v^-} := \langle \Psi_{c,C} | \widehat a^{\dagger,C}_u \widehat a^{C}_v |\Psi_{c,C} \rangle.
	\]
	We say that the CASSCF wavefunction is nondegenerate if all the eigenvalues of $\gamma_{c,C}^{\rm act}$ belong to $(0,1)$. 
	The following set of parameters
	\begin{equation}
		\overline{\mathcal{M}}:=\left\{(c,C)\in\mathbb{S}_{\mathbb{C}}^{N_{\rm det}-1}\times {U}(N_{\rm b})\mid\sigma(\gamma_{c,C}^{\rm act})\subset(0,1)\right\}
		\label{eqn:proper CASSCF state space}
	\end{equation}
	constitutes a smooth submanifold of the ambient space, where $\sigma(\gamma_{c,C}^{\rm act})$ gives the spectrum of $\gamma_{c,C}^{\rm act}$. This can be seen by noting that $\gamma^{\rm act}_{(\cdot)}$ is a smooth function in $(c, C)$ and that the set $\{\sigma(\gamma)\subset(0,1)\}$ is an open subset of $\mathbb C_{\rm herm}^{N_{\rm A}\times N_{\rm A}}$. The CASSCF manifold $\mathcal M$ is defined as the set of nondegenerate CASSCF states. In the notation of quotient manifold,\cite{absil2008optimization,boumal2023introduction} we have 
	$$\mathcal{M}:=\overline{\mathcal M}/(U(1) \times  U(N_{\rm I})\times  U(N_{\rm A})\times  U(N_{\rm E})).$$
	The constraint ``$\sigma(\gamma_{c,C}^{\rm act})\subset(0,1)$'' imposes the requirement that each of the active MOs is neither completely occupied nor unoccupied. 
	
	Let $(c,C) \in \mathbb S_{\C}^{N_{\rm det}-1} \times U(N_{\rm b})$. The tangent space $T_{[(c,C)]}\mathcal M$ to $\mathcal M$ at $[(c,C)]$ can be identified with
	\begin{widetext}
		\begin{equation}
			\left\{\left.\left(d,\underbrace{\begin{pmatrix}
					0 & -\kappa_{\rm AI}^\dagger & -\kappa_{\rm EI}^\dagger \\ 
					\kappa_{\rm AI} & 0 & -\kappa_{\rm EA}^\dagger \\  
					\kappa_{\rm EI} & \kappa_{\rm EA} & 0
			\end{pmatrix}}_{=:\kappa}\right)\in\C^{N_{\det}}\times\C^{N_{\rm b}\times N_{\rm b}}~\right|\begin{array}{l}
				d^\dagger c=0,~\kappa_{\rm AI}\in\C^{N_{\rm A}\times N_{\rm I}}\\
				\kappa_{\rm EI}\in\C^{N_{\rm E}\times N_{\rm I}},~\kappa_{\rm EA}\in\C^{N_{\rm E}\times N_{\rm A}}
			\end{array}\right\}.
			\label{eqn:CASSCF manifold tangent space}
		\end{equation}
	\end{widetext}
	In particular, $\mathcal M$ is a complex manifold of dimension 
	$$
	{\rm dim}_\C(\mathcal M) = 
	N_{\rm det}-1 + N_{\rm I} N_{\rm A} +  N_{\rm I} N_{\rm E} +  N_{\rm A} N_{\rm E} .
	$$
	Note that the formulation \eqref{eqn:CASSCF manifold tangent space} is compatible with the convention in quantum chemistry to parameterize the orbital rotation generator $\widehat{\kappa}$ as
	\begin{equation}
		\widehat{\kappa}:=\sum_{p>q}\Re(\kappa_{pq}) \widehat{E}_{pq}^-+i\sum_{p>q}\Im(\kappa_{pq}) \widehat{E}_{pq}^+\label{eqn:kappa_as}
	\end{equation}
	where $\widehat{E}^{\pm}_{pq}:=\widehat{E}_{pq}\pm\widehat{E}_{qp}:=\widehat{p}^\dagger \widehat{q} \pm \widehat{q}^\dagger \widehat{p}$ are the symmetrized and antisymmetrized singlet excitation operators. 
	
	A natural embedding of the CASSCF manifold $\mathcal M$ into the projective space $\mathcal{M}_{\rm FCI}$ is given by
	\begin{equation}
		\pi:\mathcal{M}\owns[(c,C)]\mapsto\pi([(c,C)]):=\ket{\Psi_{c,C}}\bra{\Psi_{c,C}}\in\mathcal{M}_{\rm FCI}.
		\label{eqn:embedding of CASSCF state into FCI}
	\end{equation}
	This map is the analogue for CASSCF of the Pl\"ucker embedding for HF. 
	
	The K\"ahler structure on $\mathcal M$ compatible with the many-body Schr\"odinger dynamics is the one for which $d_{[(c,C)]}\pi : T_{[(c,C)]}\mathcal M \to T_{\pi([(c,C)])} \mathcal{M}_{\rm FCI}$ is an isometry, when  $T_{\pi([(c,C)])} \mathcal{M}_{\rm FCI}$ is endowed with its natural inner product. In other words, it is defined by: for all $(d,{\kappa}), (d',{\kappa}') \in T_{[(c,C)]} \mathcal M$,
	\begin{align*}
		&~\langle (d,{\kappa}), (d',{\kappa}') \rangle_{[(c,C)]}\\
		:=&~ \langle d_{[(c,C)]}\pi (d,{\kappa}), d_{[(c,C)]}\pi (d',{\kappa}') \rangle_{\pi_{[(c,C)]}}\\
		=&~\langle d_{[(c,C)]}\pi (d,{\kappa})\mid d_{[(c,C)]}\pi (d',{\kappa}') \rangle.
	\end{align*}
	A straightforward calculation detailed in Appendix \ref{appsec:CASSCF manifold inner product} shows that
	\begin{equation}
		\begin{split}
			\langle (d,{\kappa}),&(d',{\kappa}')\rangle_{[(c,C)]} = 2\bigg(d^\dagger d'+{\rm Tr}\big((\kappa_{\rm AI}')^\top(I-\gamma_{c,C}^{\rm act})\kappa_{\rm AI}^*\big)\nonumber\\
			&+{\rm Tr}\big((\kappa_{\rm EI}')^\top\kappa_{\rm EI}^*\big)+{\rm Tr}\big(\kappa_{\rm EA}^*\gamma_{c,C}^{\rm act}(\kappa_{\rm EA}')^\top\big)\bigg).\label{eqn:CASSCF manifold inner product}
		\end{split}
	\end{equation}
	The factor of two on the right-hand side is motivated by the fact that we are working in the density-matrix space, where a variation is given by $\delta\Gamma=\ket{\delta \Psi}\bra{\Psi}+\ket{\Psi}\bra{\delta \Psi}$. Note that the inner product is positive definite due to the definition \eqref{eqn:proper CASSCF state space}. We point out that choosing a correct inner product is crucial to define the CASSCF dynamics and to study the CASSCF linear response theory (cf. the discussion at the end of Section \ref{sec:CASSCF LR}). 
	
	\par Following Part I,\cite{grazioli2026critical} the Riemannian metric and symplectic form are respectively given by
	\begin{align}
		g_{[(c,C)]}\big((d,\kappa),(d',\kappa')\big)&:=\Re\langle (d,{\kappa}),(d',{\kappa}')\rangle_{[(c,C)]},\label{eqn:CASSCF manifold Riemannian metric}\\
		\omega_{[(c,C)]}\big((d,\kappa),(d',\kappa')\big)&:=\Im\langle (d,{\kappa}),(d',{\kappa}')\rangle_{[(c,C)]},\label{eqn:CASSCF manifold symplectic form}
	\end{align}
	and the complex structure $J_{[(c,C)]}:T_{[(c,C)]}\mathcal M\to T_{[(c,C)]}\mathcal M$ satisfies $J_{[(c,C)]}^2=-{\rm Id}_{T_{[(c,C)]}\mathcal M}$. 
	The K\"ahler structure of the CASSCF manifold $\mathcal M$ is such that the three structures are compatible, in that
	\begin{equation}
		g_{[(c,C)]}\big((d,\kappa),(d',\kappa')\big)=\omega_{[(c,C)]}\bigg((d,\kappa),J_{[(c,C)]}\big((d',\kappa')\big)\bigg)
		\label{eqn:CASSCF manifold compatibility condition}
	\end{equation}
	holds for any $[(c,C)]\in\mathcal M$ and any $(d,\kappa)$, $(d',\kappa')\in T_{[(c,C)]}\mathcal M$. By the compatibility condition \eqref{eqn:CASSCF manifold compatibility condition}, we obtain the closed form of the complex structure as follows:
	\begin{equation}
		\begin{split}
			J_{[(c,C)]}&\left(\left(d,\begin{pmatrix}
				0 & -\kappa_{\rm AI}^\dagger & -\kappa_{\rm EI}^\dagger \\
				\kappa_{\rm AI} & 0 & -\kappa_{\rm EA}^\dagger \\
				\kappa_{\rm EI} & \kappa_{\rm EA} & 0
			\end{pmatrix}\right)\right):=\\
			&\left(id,\begin{pmatrix}
				0 & i\kappa_{\rm AI}^\dagger & i\kappa_{\rm EI}^\dagger \\
				i\kappa_{\rm AI} & 0 & i\kappa_{\rm EA}^\dagger \\
				i\kappa_{\rm EI} & i\kappa_{\rm EA} & 0
			\end{pmatrix}\right).
			\label{eqn:CASSCF manifold complex structure}
		\end{split}
	\end{equation}
	We remark that Eq. \eqref{eqn:CASSCF manifold compatibility condition} together with Eq. \eqref{eqn:CASSCF manifold symplectic form} imply 
	\begin{equation}
		\begin{split}
			&g_{[(c,C)]}\bigg((d,\kappa),J_{[(c,C)]}\big((d,\kappa)\big)\bigg)=\\
			&=\omega_{[(c,C)]}\bigg((d,\kappa),J_{[(c,C)]}^2\big((d,\kappa)\big)\bigg)=0,
			\label{eqn:CASSCF manifold complex structure action}
		\end{split}
	\end{equation}
	which indicates that the $J_{[(c,C)]}$ acts as a 90-degree rotation on $T_{[(c,C)]}\mathcal M$ with respect to the Riemannian metric. This point will be revisited in Section \ref{sec:TD CASSCF} when we derive the time-dependent CASSCF equations as a Hamiltonian dynamics.

	\section{Time-dependent CASSCF equations}\label{sec:TD CASSCF}
	
	Following Part I,\cite{grazioli2026critical} we need the Riemannian gradient of the energy functional on $\mathcal M$ to define the time-dependent CASSCF equations, and the Riemannian Hessian to study the linear response theory.\cite{dalgaard1980time,grazioli2026critical,helmich2019casscf,jorgensen1988linear,olsen1985linear,yeager1979multiconfigurational} We proceed by expanding the energy functional up to the second order along a smooth curve on $\mathcal{M}$ around a given state, and comparing with the Riemannian metric \eqref{eqn:CASSCF manifold Riemannian metric}.
	
	Let $[(c,C)]\in\mathcal{M}$ and $\ket{0}:=\ket{\Psi_{c,C}}$ be the associated state. For any direction $(d,\kappa)\in T_{[(c,C)]}\mathcal{M}$, consider the smooth curve $[(c(t),C(t)]$ on $\mathcal{M}$ defined as
	\begin{equation}
		c(t):=\cos(t\|d\|)c+\sin(t\|d\|)d/\|d\|,~~C(t)=Ce^{t\kappa}.
		\label{eqn:CASSCF manifold smooth curve}
	\end{equation}
	It is not difficult to verify that, at $t=0$, $(c^{(0)},C^{(0)})=(c,C)$ and the velocity $(c^{(1)},C^{(1)})$ corresponds to $(d,\kappa)\in T_{[(c,C)]}\mathcal{M}$. The associated smooth curve on the wavefunction space is 
	\begin{equation}
		\begin{split}
			&~\ket{\Psi_{c(t),C(t)}}\\
			=&~\sum_{A=1}^{N_{\det}}\big(\cos(t\|d\|)c_A+\sin(t\|d\|)d_A/\|d\|\big)e^{t\widehat\kappa}\ket{A^C},
		\end{split}
		\label{eqn:wavefunction space smooth curve}
	\end{equation}
	where $\widehat\kappa$ is the orbital rotation generator in Eq. \eqref{eqn:kappa_as}. It satisfies $\ket{\Psi^{(0)}}=\ket{0}$,
	\begin{equation}
		\ket{\Psi^{(1)}}=\sum_{A=1}^{N_{\det}}d_A\ket{A^C}+\widehat\kappa\ket{0},
		\label{eqn:wavefunction space smooth curve velocity}
	\end{equation}
	and
	\begin{equation}
		\ket{\Psi^{(2)}}=-\|d\|^2\ket{0}+2\sum_{A=1}^{N_{\det}}d_A\widehat\kappa\ket{A^C}+\widehat\kappa^2\ket{0}.
		\label{eqn:wavefunction space smooth curve acceleration}
	\end{equation}
	
	\par Let $\mathcal{E}$ and $\mathcal{E}_{\rm FCI}$ be respectively the energy functionals on $\mathcal{M}$ and $\mathcal{M}_{\rm FCI}$. They are related to each other in the sense that $\mathcal{E}([(c,C)])=\mathcal{E}_{\rm FCI}([\Psi_{c,C}])$. Therefore, their first- and second-order terms in the expansion along the curves \eqref{eqn:CASSCF manifold smooth curve} and \eqref{eqn:wavefunction space smooth curve} around $t=0$ should agree. Recall that for any smooth function $f$ over a smooth manifold $\mathcal{N}$, let $a(t)$ be a smooth curve on $\mathcal{N}$, we have the second-order Taylor expansion around $t=0$ as
	\begin{align}
		f(a(t))=&~f(a^{(0)})+t\cdot g_{a^{(0)}}\big({\rm grad}_{\mathcal{N}}f(a^{(0)}),a^{(1)}\big)\nonumber\\
		&+\frac{t^2}{2}\bigg(g_{a^{(0)}}\big({\rm Hess}_{\mathcal{N}}f(a^{(0)})[a^{(1)}],a^{(1)}\big)\nonumber\\
		&+g_{a^{(0)}}\big({\rm grad}_{\mathcal{N}}f(a^{(0)}),a^{(2)}\big)\bigg)+o(t^2).\label{eqn:second-order expansion on manifold}
	\end{align}
	Denote $\mathcal{E}_0:=\mathcal{E}([(c,C)])$. For $f=\mathcal{E}_{\rm FCI}$, $\mathcal{N}=\mathcal{M}_{\rm FCI}$, we have from Eq. \eqref{eqn:wavefunction space smooth curve velocity} that
	\begin{equation}
		{\rm grad}_{\mathcal{M}_{\rm FCI}}\mathcal{E}_{\rm FCI}([\Psi^{(0)}])=(\widehat H-\mathcal{E}_0)\ket{0}\bra{0}+\text{h.c.}\label{eqn:FCI energy Riemannian gradient}
	\end{equation}
	and
	\begin{align}
		&~{\rm Hess}_{\mathcal{M}_{\rm FCI}}\mathcal{E}_{\rm FCI}([\Psi^{(0)}])[\Psi^{(1)}]\label{eqn:FCI energy Riemannian Hessian}\\
		=&~(I-\ket{0}\bra{0})(\widehat H-\mathcal{E}_0)\bigg(\sum_{A=1}^{N_{\det}}d_A\ket{A^C}+\widehat\kappa\ket{0}\bigg)\bra{0}+\text{h.c.}\nonumber
	\end{align}
	Plugging Eqs. \eqref{eqn:FCI energy Riemannian gradient} and \eqref{eqn:FCI energy Riemannian Hessian} into Eq. \eqref{eqn:second-order expansion on manifold} and by Eqs. \eqref{eqn:wavefunction space smooth curve velocity} and \eqref{eqn:wavefunction space smooth curve acceleration}, straightforward calculations give
	\begin{align}
		&~\mathcal{E}_{\rm FCI}([\Psi_{c(t),C(t)}])=\mathcal{E}_0\nonumber\\
		&+2t\cdot\Re\bigg(\sum_{A=1}^{N_{\det}}d_A\braket{0|(\widehat H-\mathcal{E}_0)|A^C}+\braket{0|\widehat H\widehat\kappa|0}\bigg)\nonumber\\
		&+t^2\cdot\Re\bigg(\sum_{A,A'=1}^{N_{\det}}d_A^*d_{A'}\braket{A^C|(\widehat H-\mathcal{E}_0)|A^{'C}}      \label{eqn:FCI energy second-order expansion}\\
		&+2\sum_{A=1}^{N_{\det}}d_A\braket{0|[\widehat H,\widehat\kappa]|A^C}+\frac12\braket{0|[[\widehat H,\widehat\kappa],\widehat\kappa]|0}\bigg)+o(t^2).\nonumber
	\end{align}
	On the other hand, by Eq. \eqref{eqn:second-order expansion on manifold},
	\begin{align}
		&~\mathcal{E}([c(t),C(t)])=\mathcal{E}_0+t\cdot g_{[(c,C)]}\big((d^{\rm grad},\kappa^{\rm grad}),(d,\kappa)\big)\nonumber\\
		&+\frac{t^2}{2}\bigg(g_{[(c,C)]}\big((d^{\rm Hess},\kappa^{\rm Hess}),(d,\kappa)\big)\label{eqn:CASSCF energy second-order expansion}\\
		&+g_{[(c,C)]}\big((d^{\rm grad},\kappa^{\rm grad}),(d^{\rm proj},\kappa^{\rm proj})\big)\bigg)+o(t^2).\nonumber
	\end{align}
	Here, $(d^{\rm grad},\kappa^{\rm grad}),(d^{\rm Hess},\kappa^{\rm Hess})\in T_{[(c,C)]}\mathcal{M}$ are respectively the representations of the Riemannian gradient of $\mathcal E$ at $[(c,C)]$ and the action of the Riemannian Hessian of $\mathcal E$ at $[(c,C)]$ on $(d,\kappa)$, $(d^{\rm proj},\kappa^{\rm proj})\in T_{[(c,C)]}\mathcal{M}$ is the representation of the projection of $\Psi^{(2)}$ onto the tangent space $T_{[\Psi_{c,C}]}\pi(\mathcal M)$ (which is not necessarily zero because the smooth curve $[(c(t),C(t)]$ is not a geodesics on $\mathcal M$ in general). 
	
	\par Equating the first-order terms in Eqs. \eqref{eqn:FCI energy second-order expansion} and \eqref{eqn:CASSCF energy second-order expansion}, we obtain
	\begin{equation*}
		\begin{split}
			&g_{[(c,C)]}\big((d^{\rm grad},\kappa^{\rm grad}),(d,\kappa)\big)=
			\\ &=2\Re\bigg(\sum_{A=1}^{N_{\det}}d_A\braket{0|(\widehat H-\mathcal{E}_0)|A^C}+\braket{0|\widehat H\widehat\kappa|0}\bigg).\end{split}
	\end{equation*}
	By comparing with Eq. \eqref{eqn:CASSCF manifold Riemannian metric} and using the arbitrariness of $(d,\kappa)$, we obtain 
	\begin{align}
		d^{\rm grad}&
		=\big(\braket{A^C|(\widehat H-\mathcal E_0)|0}\big)_{A=1}^{N_{\rm det}},\label{eqn:CASSCF energy Riemannian gradient d}\\
		\kappa_{\rm AI}^{\rm grad}&=(I-\gamma_{c,C}^{\rm act})^{-\top}\big(\braket{0|\widehat i^\dagger\widehat u\widehat H|0}\big)_{u^-,i},\label{eqn:CASSCF energy Riemannian gradient k_AI}\\
		\kappa_{\rm EI}^{\rm grad}&=\big(\braket{0|\widehat i^\dagger\widehat a\widehat H|0}\big)_{a^-,i},\label{eqn:CASSCF energy Riemannian gradient k_EI}\\
		\kappa_{\rm EA}^{\rm grad}&=\big(\braket{0|\widehat u^\dagger\widehat a\widehat H|0}\big)_{a^-,u^-}(\gamma_{c,C}^{\rm act})^{-\top}.\label{eqn:CASSCF energy Riemannian gradient k_EA}
	\end{align}
	Here $a^-:=a-N_{\rm I}-N_{\rm A}$ denotes similarly the shifted index for the $a$-th MO, which is in fact the $a^-$-th external MO ($a=N_{\rm I}+N_{\rm A}+1,\ldots,N_{\rm b}$). 
	
	\par The time-dependent CASSCF equation is the following Hamiltonian dynamics on $\mathcal M$:
	\begin{equation}
		\frac{{\rm d}[(c(t),C(t))]}{{\rm d}t}=J_{[(c(t),C(t))]}\bigg({\rm grad}_{\mathcal{M}}\mathcal{E}\big([(c(t),C(t))]\big)\bigg).
		\label{eqn:CASSCF Hamiltonian dynamics}
	\end{equation}
	From Eq. \eqref{eqn:CASSCF manifold complex structure action}, the dynamics \eqref{eqn:CASSCF Hamiltonian dynamics} can be understood as a ``rotated'' gradient flow, which conserves the Hamiltonian (i.e., the CASSCF energy) along the trajectory. 
	
	
	\par Let us turn to the Riemannian Hessian-vector product. Since the second term in the big parentheses of Eq. \eqref{eqn:CASSCF energy second-order expansion} is difficult to compute but is in fact a quantity determined by $(d^{\rm grad},\kappa^{\rm grad})$, we assume that $[(c,C)]$ is a critical point of the CASSCF energy $\mathcal{E}$ on $\mathcal M$. As a result, equating the second-order terms in Eqs. \eqref{eqn:FCI energy second-order expansion} and \eqref{eqn:CASSCF energy second-order expansion} gives
	\begin{align*}
		&g_{[(c,C)]}\big((d^{\rm Hess},\kappa^{\rm Hess}),(d,\kappa)\big)\\&=2\Re\bigg(\sum_{A,A'=1}^{N_{\det}}d_A^*d_{A'}\braket{A^C|(\widehat H-\mathcal{E}_0)|A^{'C}}\\
		&+2\sum_{A=1}^{N_{\det}}d_A\braket{0|[\widehat H,\widehat\kappa]|A^C}+\frac12\braket{0|[[\widehat H,\widehat\kappa],\widehat\kappa]|0}\bigg).
	\end{align*}
	Again, by comparing with Eq. \eqref{eqn:CASSCF manifold Riemannian metric} and using the arbitrariness of $(d,\kappa)$, we obtain 
	\begin{widetext}
		\begin{align}
			d^{\rm Hess}&=\bigg(\sum_{A'=1}^{N_{\det}}d_{A'}\braket{A^C|(\widehat H-\mathcal{E}_0)|A^{'C}}+\braket{A^C|[\widehat H,\widehat\kappa]|0}\bigg)_{A=1}^{N_{\rm det}},\label{eqn:CASSCF energy Riemannian Hessian d}\\
			\kappa_{\rm AI}^{\rm Hess}&=(I-\gamma_{c,C}^{\rm act})^{-\top}\bigg(\sum_{A=1}^{N_{\det}}\big(d_A\braket{0|[\widehat i^\dagger\widehat u,\widehat H]|A^C}+d_A^*\braket{A^C|[\widehat i^\dagger\widehat u,\widehat H]|0}\big)\nonumber\\
			&\qquad+\sum_{r>s}\big(\kappa_{rs}^*\braket{0|[[\widehat H,\widehat i^\dagger\widehat u],\widehat s^\dagger\widehat r]|0}-\kappa_{rs}\braket{0|[[\widehat{H},\widehat i^\dagger\widehat u],\widehat r^\dagger\widehat s]|0}\big)\bigg)_{u^-,i},\label{eqn:CASSCF energy Riemannian Hessian k AI}\\
			\kappa_{\rm EI}^{\rm Hess}&=\bigg(\sum_{A=1}^{N_{\det}}\big(d_A\braket{0|[\widehat i^\dagger\widehat a,\widehat H]|A^C}+d_A^*\braket{A^C|[\widehat i^\dagger\widehat a,\widehat H]|0}\big)\nonumber\\
			&\qquad+\sum_{r>s}\big(\kappa_{rs}^*\braket{0|[[\widehat H,\widehat i^\dagger\widehat a],\widehat s^\dagger\widehat r]|0}-\kappa_{rs}\braket{0|[[\widehat{H},\widehat i^\dagger\widehat a],\widehat r^\dagger\widehat s]|0}\big)\bigg)_{a^-,i},\label{eqn:CASSCF energy Riemannian Hessian k EI}\\
			\kappa_{\rm EA}^{\rm Hess}&=\bigg(\sum_{A=1}^{N_{\det}}\big(d_A\braket{0|[\widehat u^\dagger\widehat a,\widehat H]|A^C}+d_A^*\braket{A^C|[\widehat u^\dagger\widehat a,\widehat H]|0}\big)\nonumber\\
			&\qquad+\sum_{r>s}\big(\kappa_{rs}^*\braket{0|[[\widehat H,\widehat u^\dagger\widehat a],\widehat s^\dagger\widehat r]|0}-\kappa_{rs}\braket{0|[[\widehat{H},\widehat u^\dagger\widehat a],\widehat r^\dagger\widehat s]|0}\big)\bigg)_{u^-,a^-}(\gamma_{c,C}^{\rm act})^{-\top}.\label{eqn:CASSCF energy Riemannian Hessian k EA}
		\end{align}
	\end{widetext}

	\section{CASSCF linear response theory}\label{sec:CASSCF LR}
	
	The Hamiltonian dynamics \eqref{eqn:CASSCF Hamiltonian dynamics} can be linearized at a stable equilibrium state $[(c_\star,C_\star)]$ as
	\begin{multline}
		\frac{{\rm d}(d(t),\kappa(t))}{{\rm d}t}=\\J_{[(c_\star,C_\star)]}\bigg({\rm Hess}_{\mathcal M}\mathcal E\big([(c_\star,C_\star)]\big)[(d(t),\kappa(t))]\bigg),
		\label{eqn:CASSCF linearized Hamiltonian dynamics}
	\end{multline}
	where $(d(t),\kappa(t))\in T_{[(c_\star,C_\star]}\mathcal{M}$. Here, the stability of $[(c_\star,C_\star)]$ means that ${\rm Hess}_{\mathcal M}\mathcal E([(c_\star,C_\star)])$ is positive definite on the tangent space $T_{[(c_\star,C_\star)]}\mathcal M$. In the following, we show that the linear response theory for CASSCF consists in studying the vibration frequencies of the linearized dynamics \eqref{eqn:CASSCF linearized Hamiltonian dynamics} on $T_{[(c_\star,C_\star)]}\mathcal M$, which are identical to the symplectic eigenvalues of ${\rm Hess}_{\mathcal M}\mathcal E([(c_\star,C_\star)])$ on $T_{[(c_\star,C_\star)]}\mathcal M$,\cite{grazioli2026critical} thereby providing a systematic and straightforward way to derive the CASSCF linear response equations.
	
	For this purpose, we need to identify a canonical orthonormal basis of $T_{[(c_\star,C_\star)]}\mathcal M$. It is convenient to work with a unitary matrix $C_\star$ for which the matrix $\gamma^{\rm act}_{c_\star,C_\star}$ is diagonal (the CI vector $c_\star$ should be transformed accordingly by Eq. \eqref{eqn:quotient structure transformation on c}):
	\begin{equation*}
		\begin{split}
			\gamma^{\rm act}_{c_\star,C_\star} &= {\rm diag}(n_{N_{\rm I}+1}, \cdots , n_{N_{\rm I}+N_{\rm A}}),\\
			&\quad 1 > n_{N_{\rm I}+1} \ge \cdots \ge n_{N_{\rm I}+N_{\rm A}} > 0,
		\end{split}
	\end{equation*}
	where the $n_u$'s are the natural occupation numbers in the active space. By the inner product \eqref{eqn:CASSCF manifold inner product}, a canonical real orthonormal basis of $T_{[(c_\star,C_\star)]}\mathcal M$ is then
	\begin{align}
		&\mathcal B_{\R}:=\label{eqn:CASSCF manifold real orthonormal basis}\\
		&\bigg(\big\{(e_k,0)\big\}_{k=1}^{N_{\rm det}-1};\big\{(0,E_{ui})\big\}_{u,i};\big\{(0,E_{ai})\big\}_{a,i}; \big\{(0,E_{au})\big\}_{a,u};\nonumber\\
		&\big\{-J_{[(c_\star,C_\star)]}\big((e_k,0)\big)\big\}_{k=1}^{N_{\rm det}-1};\big\{-J_{[(c_\star,C_\star)]}\big((0,E_{ui})\big)\big\}_{u,i};\nonumber\\
		&\big\{-J_{[(c_\star,C_\star)]}\big((0,E_{ai})\big)\big\}_{a,i}; \big\{-J_{[(c_\star,C_\star)]}\big((0,E_{au})\big)\big\}_{a,u}\bigg),\nonumber
	\end{align}
	where $(e_k)_{1 \le k \le N_{\rm det}-1}$ form a canonical basis (scaled by $1/\sqrt{2}$) of the hyperplane orthogonal to $c_\star$ in $\C^{N_{\det}}$, 
	\begin{equation*}
		\begin{split}
			&[E_{ui}]_{pq}:=\frac{t_{ui}}{\sqrt{2}}(\delta_{pu}\delta_{qi}-\delta_{pi}\delta_{qu}),\\
			&
			[E_{ai}]_{pq}:=\frac{t_{ai}}{\sqrt{2}}(\delta_{pa}\delta_{qi}-\delta_{pi}\delta_{qa}),\\
			&
			[E_{au}]_{pq}:=\frac{t_{au}}{\sqrt{2}}(\delta_{pa}\delta_{qu}-\delta_{pu}\delta_{qa}),
		\end{split}
	\end{equation*}
	with
	$$t_{ui}:=\frac{1}{\sqrt{1-n_u}},\qquad t_{ai}:=1,\qquad t_{au}:=\frac{1}{\sqrt{n_u}}.$$
	
	\par Note that closed forms of the basis vectors $\{(e_k,0)\}$ remain unclear, which would render the matrix representation of the Riemannian Hessian less explicit. Therefore, we further augment the basis \eqref{eqn:CASSCF manifold real orthonormal basis} to
	\begin{align}
		&\mathcal B_{\R}^{\rm aug}:=\label{eqn:CASSCF manifold real orthonormal basis augment}\\
		&\bigg(\big\{(\tilde e_A,0)\big\}_{A=1}^{N_{\rm det}};\big\{(0,E_{ui})\big\}_{u,i};\big\{(0,E_{ai})\big\}_{a,i}; \big\{(0,E_{au})\big\}_{a,u};\nonumber\\
		& \big\{(-i\tilde e_A,0)\big\}_{A=1}^{N_{\rm det}};\big\{-J_{[(c_\star,C_\star)]}\big((0,E_{ui})\big)\big\}_{u,i};\nonumber\\
		& \big\{-J_{[(c_\star,C_\star)]}\big((0,E_{ai})\big)\big\}_{a,i}; \big\{-J_{[(c_\star,C_\star)]}\big((0,E_{au})\big)\big\}_{a,u}\bigg).\nonumber
	\end{align}
	The vectors $\{\tilde e_A\}_{A=1}^{N_{\det}}$ constitute the canonical basis of $\C^{N_{\det}}$ (scaled by $1/\sqrt{2}$). The augmented set \eqref{eqn:CASSCF manifold real orthonormal basis augment} will no longer serve to be a basis set of the tangent space since the orthogonality constraint is not enforced. As a result, the Riemannian Hessian at $[(c_\star,C_\star)]$ needs to be smoothly extended to vectors outside $T_{[(c_\star,C_\star)]}\mathcal M$. Nevertheless, this redundancy is almost harmless: the representation of the (extended) Riemannian Hessian under the augmented set \eqref{eqn:CASSCF manifold real orthonormal basis augment} shares the non-zero part of the spectrum with that under the basis \eqref{eqn:CASSCF manifold real orthonormal basis}, at the only price of introducing an extra zero eigenvalue. 
	
	\par The augmented set \eqref{eqn:CASSCF manifold real orthonormal basis augment} also corresponds to the conventional basis adopted in quantum chemistry. The basis vector $(\tilde e_A,0)$ corresponds to the single Slater determinant $\frac{1}{\sqrt{2}}\ket{A^C}$, while $(-i\tilde e_A,0)$ to $-\frac{i}{\sqrt{2}}\ket{A^C}$. The basis vector $(0,E_{pq})$ corresponds to the excited state $\frac{t_{pq}}{\sqrt{2}}\widehat E_{pq}^-\ket{\Psi}$, while $-J_{[(c_\star,C_\star)]}\big((0,E_{pq})\big)$ to the excited state $-i\frac{t_{pq}}{\sqrt{2}}\widehat E_{pq}^+\ket{\Psi}$. 

	\par Next, we form the matrix representation of the extended Riemannian Hessian (denoted by ${\rm H}_\star$) under the augmented set \eqref{eqn:CASSCF manifold real orthonormal basis augment}:
	$$\mathfrak{H}_\star:=\begin{pmatrix}
		\mathfrak{H}_\star^{\rm rr} & \mathfrak{H}_\star^{\rm ri}\\
		\mathfrak{H}_\star^{\rm ir} & \mathfrak{H}_\star^{\rm ii}
	\end{pmatrix},$$
	where, for $A,A'=1,\ldots,N_{\det}$ and $p>q$, $r>s$ in different groups,
	\begin{align*}
		[\mathfrak{H}_\star^{\rm rr}]_{A,A'}:=&~g_{[(c_\star,C_\star)]}\big((\tilde e_A,0),{\rm H}_\star[(\tilde e_{A'},0)]\big)\\
		=&~\Re\braket{A^C|(\widehat H-\mathcal E_0)|A^{'C}},\\
		[\mathfrak{H}_\star^{\rm rr}]_{A,pq}:=&~g_{[(c_\star,C_\star)]}\big((\tilde e_A,0),{\rm H}_\star[(0,E_{pq})]\big)\\
		=&~t_{pq}\cdot\Re\braket{A^C|[\widehat H,\widehat E_{pq}^-]|0},\\
		[\mathfrak{H}_\star^{\rm rr}]_{pq,A}:=&~g_{[(c_\star,C_\star)]}\big((0,E_{pq}),{\rm H}_\star[(\tilde e_A,0)]\big)\\
		=&~t_{pq}\cdot\Re\braket{0|[\widehat E_{qp}^-,\widehat H]|A^C},\\
		[\mathfrak{H}_\star^{\rm rr}]_{pq,rs}:=&~g_{[(c_\star,C_\star)]}\big((0,E_{pq}),{\rm H}_\star[(0,E_{rs})]\big)\\
		=&~\frac{t_{pq}t_{rs}}{2}\cdot\Re\braket{0|[[\widehat H,\widehat E_{pq}^-],\widehat E_{rs}^-]|0}.
	\end{align*}
	The submatrices $\mathfrak{H}_\star^{\rm ri}$, $\mathfrak{H}_\star^{\rm ir}$, and $\mathfrak{H}_\star^{\rm ii}$ are defined in a similar way. In particular, for $\mathfrak{H}_\star^{\rm ii}$, 
	\begin{align*}
		[\mathfrak{H}_\star^{\rm ii}]_{A,A'}&:=\Re\braket{A^C|(\widehat H-\mathcal E_0)|A^{'C}},\\
		[\mathfrak{H}_\star^{\rm ii}]_{A,pq}&:=t_{pq}\cdot\Re\braket{A^C|[\widehat H,\widehat E_{pq}^+]|0},\\
		[\mathfrak{H}_\star^{\rm ii}]_{pq,A}&:=t_{pq}\cdot\Re\braket{0|[\widehat E_{pq}^+,\widehat H]|A^C},\\
		[\mathfrak{H}_\star^{\rm ii}]_{pq,rs}&:=-\frac{t_{pq}t_{rs}}{2}\cdot\Re\braket{0|[[\widehat H,\widehat E_{pq}^+],\widehat E_{rs}^+]|0}.
	\end{align*}
	From the definition of the Riemannian metric $g_{[(c_\star,C_\star)]}$ in Eq. \eqref{eqn:CASSCF manifold Riemannian metric} and complex structure in Eq. \eqref{eqn:CASSCF manifold complex structure}, $\mathfrak{H}_\star^{\rm ri}$ and $\mathfrak{H}_\star^{\rm ir}$ vanish and $\mathfrak{H}_\star$ becomes block diagonal. Following Part I,\cite{grazioli2026critical} the symplectic eigenvalues of $\mathfrak{H}_\star$ are thus identical to the eigenvalues of the following half-sized matrix:
	\begin{equation}
		\big((\mathfrak{H}_\star^{\rm rr})^{1/2}\mathfrak{H}_\star^{\rm ii}(\mathfrak{H}_\star^{\rm rr})^{1/2}\big)^{1/2}.
		\label{eqn:CASSCF LR reduced size}
	\end{equation}
	
	\par The above derivations agree with the CASSCF linear response equations in quantum chemistry.\cite{olsen1985linear} In the usual procedure for CASSCF, the excitation energies are found by computing the generalized eigenvalues of the matrix pair
	\begin{equation}
		\left(\begin{pmatrix}
			\mathbf{A+B} &\mathbf{0}\\\mathbf{0} & \mathbf{A-B}
		\end{pmatrix},\begin{pmatrix}
			\mathbf{0} & \mathbf{\Sigma} \\
			\mathbf{\Sigma} &\mathbf{0}
		\end{pmatrix}\right),\label{eqn:LR-CASSCF}
	\end{equation}
	where the submatrices are defined as (see, e.g., Nottoli \textit{et al.}\cite{nottoli2025efficient})
	\begin{align*}
		[\mathbf{A\pm B}]_{A,A'}&:=\braket{A^C|(\widehat{H}-\mathcal E_0)|A^{'C}},\\
		[\mathbf{A\pm B}]_{A,pq}&:=\braket{0|[\widehat E_{pq}^{\mp},\widehat{H}]|A^C},\\
		[\mathbf{A\pm B}]_{pq,rs}&:=\pm\frac12\braket{0|[[\widehat{H},\widehat E_{pq}^{\mp}],\widehat E_{rs}^{\mp}]|0},
	\end{align*}
	and
	$$[\mathbf{\Sigma}]_{A,A'}:=\delta_{A,A'},\quad[\mathbf{\Sigma}]_{pq,rs}:=\braket{0|[\widehat q^\dagger\widehat p,\widehat r^\dagger\widehat s]|0}.$$
	After a transformation of variables, the generalized eigenvalues of the matrix pair \eqref{eqn:LR-CASSCF} can be shown to be identical to those of
	$$\left(\begin{pmatrix}
		\tilde{\mathbf{A}}+\tilde{\mathbf{B}} & \mathbf 0\\
		\mathbf 0 & \tilde{\mathbf{A}}-\tilde{\mathbf{B}}
	\end{pmatrix},\begin{pmatrix}
		\mathbf 0 & \mathbf I\\
		\mathbf I & \mathbf 0
	\end{pmatrix}\right)$$
	with $\tilde{\mathbf{A}}:=\mathbf{\Sigma}^{-\frac12}\mathbf{A}\mathbf{\Sigma}^{-\frac12}$ and $\tilde{\mathbf{B}}:=\mathbf{\Sigma}^{-\frac12}\mathbf{B}\mathbf{\Sigma}^{-\frac12}$, which are in turn identical to the eigenvalues of 
	\begin{equation}
		\big((\tilde{\mathbf{A}}+\tilde{\mathbf{B}})^{1/2}(\tilde{\mathbf{A}}-\tilde{\mathbf{B}})(\tilde{\mathbf{A}}+\tilde{\mathbf{B}})^{1/2}\big)^{1/2}.
		\label{eqn:LR_CASSCF transform}
	\end{equation}
	We conclude that Eqs. \eqref{eqn:LR_CASSCF transform} and \eqref{eqn:CASSCF LR reduced size} coincide after noting $\mathfrak{H}_\star^{\rm rr}=\tilde{\mathbf{A}}+\tilde{\mathbf{B}}$ and $\mathfrak{H}_\star^{\rm ii}=\tilde{\mathbf{A}}-\tilde{\mathbf{B}}$. 
	
	\par We end this section by pointing out that defining the K\"ahler structure on the CASSCF manifold $\mathcal M$ by using the isometric embedding \eqref{eqn:embedding of CASSCF state into FCI} is crucial to obtain the correct results. Indeed, we use the inner product to define $g_{[(c,C)]}$, $\omega_{[(c,C)]}$, and $J_{[(c,C)]}$, then use $g_{[(c,C)]}$ to define the Riemannian gradient and Hessian, and finally, on top of these, define the CASSCF dynamics and study its linearization. 
	
	\section{CASSCF ground and excited states}\label{sec:CASSCF SS}
	
	Any critical point $[(c_\star,C_\star)]$ of the CASSCF energy functional is the stationary solution to the CASSCF dynamics \eqref{eqn:CASSCF Hamiltonian dynamics} and solves the time-independent equation
	\begin{equation*}
		{\rm grad}_{\mathcal{M}}\mathcal{E}([(c_\star,C_\star)])=0,
	\end{equation*}
	which, by Eqs. \eqref{eqn:CASSCF energy Riemannian gradient d}-\eqref{eqn:CASSCF energy Riemannian gradient k_EA}, is equivalent to 
	\begin{align}
		&\braket{A^{C_\star}|(\widehat H-\mathcal{E}_0)|0}=0,\quad\forall~A=1,\ldots,N_{\det};\label{eqn:CASSCF equation d}\\
		&\braket{0|\widehat H\widehat p^\dagger\widehat q|0}=0,\quad\forall~p>q~\text{in different groups}.\label{eqn:CASSCF equation k}
	\end{align}
	These recover Eqs. (15) and (16) in Ref. \citenum{jensen1984direct}. 
	
	\par The CASSCF ground state is the critical point with the lowest CASSCF energy. The CASSCF ground-state minimization can be cast as
	$$\min_{[(c,C)]\in\mathcal M}~\mathcal{E}([(c,C)]).$$
	There have been well-established algorithms for this purpose, such as super CI\cite{angeli2002novel,kollmar2019perturbation,malmqvist1990restricted,roos1980complete,roos1980complete2,siegbahn1981complete} and the Norm Extended Optimization (NEO).\cite{jensen1986direct,jensen1984direct} Moreover, note that the minimization with respect to $[c]$ can be further wrapped up as an inner problem over a high-dimensional sphere. Consequently, the outer problem involving $[C]$ can be formulated on the flag manifold ${\rm Flag}(N_{\rm I},N_{\rm I}+N_{\rm A};N_{\rm b})$.\cite{ye2022optimization} This structure has recently been made use of to develop fast convergent ground-state minimization algorithms that only require first-order information.\cite{vidal2024geometric} 
	
	\par The CASSCF excited states are defined as saddle points of the CASSCF energy landscape, where the Riemannian gradient vanishes (cf. the CASSCF equations \eqref{eqn:CASSCF equation d} and \eqref{eqn:CASSCF equation k}) and the Riemannian Hessian has negative eigenvalues. If the CASSCF energy functional $\mathcal{E}$ is a Morse function on $\mathcal{M}$, i.e., ${\rm Hess}_{\mathcal{M}}\mathcal{E}$ is nondegenerate at any critical point,\cite{milnor1963morse} the number of negative eigenvalues of ${\rm Hess}_{\mathcal{M}}\mathcal{E}$ at a saddle point is called its Morse index. 
	
	\par Approximating the excited states in exact theory with the CASSCF excited states underlies the state-specific methods for excited-state calculations.\cite{hanscam2022applying,helmich2022trust,jensen1986direct,jensen1984direct,tran2020improving,tran2019tracking} 
	Nevertheless, since the CASSCF theory introduces nonlinear approximations, the indices of CASSCF excited states do not necessarily correspond to the orders of excitations in exact theory; moreover, each excited state in exact theory could possibly be approximated by multiple CASSCF excited states (with different indices), and spurious CASSCF excited states that do not possess any physical meaning can emerge (cf. Fig. 1 in Ref. \citenum{grazioli2026critical}). 
	
	\par Setting aside these fundamental issues, searching for the CASSCF excited states already poses significant challenges. Existing state-specific methods can face convergence difficulties (e.g., root flipping and variational collapse) due to the intrinsic instability of saddle points, introduce non-critical solutions,\cite{hanscam2022applying,tran2020improving,tran2019tracking} or explicitly involve the computationally expensive Riemannian Hessian.\cite{helmich2022trust,jensen1986direct,jensen1984direct} 
	
	
	\par Following the developments in the unconstrained settings\cite{barkema1996event,cances2009some,e2011gentlest,henkelman1999dimer,zhang2012shrinking} and the recent advances in the constrained settings,\cite{hu2026constrained,yin2022constrained} and based on the developments in Sections \ref{sec:CASSCF manifold Kahler} and \ref{sec:TD CASSCF}, we propose a first-order Constrained Gentlest Ascent Method (CGAM) for computing the CASSCF excited states of any index, which proceeds in each iteration as follows: with the current state $[(c^{(t)},C^{(t)})]\in\mathcal{M}$ and current estimates of lowest eigenvectors $\{(d_i^{(t)},\kappa_i^{(t)})\}_{i=1}^k\subseteq T_{[(c^{(t)},C^{(t)})]}\mathcal M$, where $k\in\mathbb{N}$ denotes the target index of CASSCF excited states, and the step size $\eta>0$.
	
	\medskip
	
	\par\noindent\textbf{S1.} Compute the search directions for the state 
	\begin{align*}
		\big(\bm\delta_c^{(t)},\bm\delta_C^{(t)}\big):=&-\big(d^{{\rm grad},(t)},\kappa^{{\rm grad},(t)}\big)+2\sum_{i=1}^k\alpha_i^{(t)}\big(d_i^{(t)},\kappa_i^{(t)}\big),
	\end{align*}
	where $\big(d^{{\rm grad},(t)},\kappa^{{\rm grad},(t)}\big)\in T_{[(c^{(t)},C^{(t)})]}\mathcal M$ denotes the representation of the Riemannian gradient at $[(c^{(t)},C^{(t)})]$ (cf. Eqs. \eqref{eqn:CASSCF energy Riemannian gradient d}-\eqref{eqn:CASSCF energy Riemannian gradient k_EA}),
	$$\alpha_i^{(t)}:=g_{[(c^{(t)},C^{(t)})]}\big((d_i^{(t)},\kappa_i^{(t)}),(d^{{\rm grad},(t)},\kappa^{{\rm grad},(t)})\big).$$
	
	\smallskip 
	
	\par\noindent\textbf{S2.} Compute the search directions for the estimates of lowest eigenvectors
	\begin{align*}\big(\bm\delta_{d_i}^{(t)},\bm\delta_{\kappa_i}^{(t)}\big):=&-\big(d_i^{{\rm Hess},(t)},\kappa_i^{{\rm Hess},(t)}\big)+\sum_{i=1}^k\beta_i^{(t)}\big(d_i^{(t)},\kappa_i^{(t)}\big),\end{align*}
	where $\big(d_i^{{\rm Hess},(t)},\kappa_i^{{\rm Hess},(t)}\big)\in T_{[(c^{(t)},C^{(t)})]}\mathcal M$ denotes the representation of the {\it approximation} to the Riemannian Hessian at $[(c^{(t)},C^{(t)})]$ acting on $[(d_i^{(t)},\kappa_i^{(t)})]$ (cf. Eqs. \eqref{eqn:CASSCF energy Riemannian Hessian d}-\eqref{eqn:CASSCF energy Riemannian Hessian k EA}), 
	$$\beta_i^{(t)}:=g_{[(c^{(t)},C^{(t)})]}\big((d_i^{(t)},\kappa_i^{(t)}),(d_i^{{\rm Hess},(t)},\kappa_i^{{\rm Hess},(t)})\big).$$
	
	\smallskip 
	
	\par\noindent\textbf{S3.} Update the state
	$$c^{(t+1)}:=\frac{c^{(t)}+\eta\bm\delta_c^{(t)}}{\|c^{(t)}+\eta\bm\delta_c^{(t)}\|},~~ C^{(t+1)}:={\rm orth}\big(C^{(t)}(I+\eta\bm\delta_C^{(t)})\big),$$
	where 
	``orth'' refers to any orthonormalization process such as QR decomposition. 
	
	\smallskip 
	
	\par\noindent\textbf{S4.} Update the estimates of lowest eigenvectors
	\begin{align*}\big(d_i^{(t+1)},\kappa_i^{(t+1)}\big):=&\big(d_i^{(t)},\kappa_i^{(t)}\big)+\eta\big(\bm\delta_{d_i}^{(t)},\bm\delta_{\kappa_i}^{(t)}\big).
	\end{align*}
	
	\medskip
	
	\par The idea underlying the CGAM is as follows: instead of performing Riemannian gradient descent to local minima, it increases the energy by climbing up along $\{\pm(d_i^{(t)},\kappa_i^{(t)})\}_{i=1}^k$ (whose signs are determined by their angles with the Riemannian gradient), while decreases the energy in all the perpendicular directions (cf. S1 and S3). In other words, the Riemannian gradient is \textit{reflected} so that the saddle points become locally stable. The subspace spanned by $\{(d_i^{(t)},\kappa_i^{(t)})\}_{i=1}^k$ approximates the lowest $k$-dimensional invariant subspace of the Riemannian Hessian at $[(c^{(t)},C^{(t)})]$, by following the gradient flow of the associated Rayleigh-Ritz minimization (cf. S2 and S4). Note that in S2, the CGAM requires at most Riemannian Hessian-vector products, which could be further approximated through finite-difference schemes (cf. dimer methods in the literature\cite{henkelman1999dimer,zhang2012shrinking}), thereby completely eliminating the use of second-order information. In all, compared with existing state-specific methods, the CGAM is more stable and more computationally economic per iteration, and allows users to specify the target index of CASSCF excited states. By slight modifications to the analysis in Ref. \citenum{hu2026constrained}, we are able to establish its local linear convergence. 
	
	
	\par With nice intuitions and convergence theories though, the CGAM is in general not suitable for computing CASSCF excited states up to high accuracy. Its numerical performance as well as theoretical local convergence rates have been found to be sensitive to condition numbers, i.e., the ratio of maximum absolute eigenvalue over minimum absolute eigenvalue of the Riemannian Hessian.\cite{hu2026constrained} This deficiency is not surprising, as the CGAM is only first-order in terms of both $[(c,C)]$ and $\{(d_i,\kappa_i)\}_{i=1}^k$. As consequence, we could anticipate its potential poor efficiency on ill-conditioned energy landscapes. Nevertheless, the robustness of the CGAM should stand out as a significant advantage. 
	
	\par The CGAM could be tremendously accelerated locally, without explicitly involving second-order oracles, if the first-order information is accumulated in a proper way. Indeed, we could construct locally strongly convex merit functions along the trajectory, which enable limited-memory Broyden-Fletcher-Goldfarb-Shanno (BFGS) quasi-Newton approximations.\cite{liu1989limited} The quasi-Newton strategy will be detailed in a separate work. By combining the CGAM with the quasi-Newton updates, we are able to robustly navigate through the CASSCF energy landscape, while achieving local superlinear convergence.

	\section{Numerical experiments}\label{sec:numerics}
	
	In this section, we apply the CGAM method to compute the excited states of three simple molecules, namely, water with CAS(8,6), formaldehyde with CAS(4,3), and ethylene with CAS(2,2). We choose such simple systems because they are computationally accessible, allowing us to perform a large number of numerical experiments, and because their excited states are well known and easily identified. 
	The geometries of the examined molecules can be found in the supplementary material. The calculations have been performed using the 6-31G basis set.\cite{binkley1980self} All states discussed in this paper are singlet states. We first perform state-specific calculations, using both the CGAM coupled to the CFOUR program and the NEO algorithm within the \textsc{Dalton} program. 
	The obtained states are analyzed in order to identify their physical interpretations of excited states.
	We then compare the state-specific results with those given by state-averaged and linear response calculations. 
	
	\medskip
	
	\par\noindent\textbf{Computational details.} For the state-specific calculations, we target the index-1 and index-2 saddle points of the energy landscapes. 
	For each Morse index, we start the CGAM with either chemically motivated initializations -- CASCI ground state (``gs''), first (``1ex'') and second (``2ex'') excited states -- or random initializations (500 trials for each case). 
	When using the NEO algorithm within the \textsc{Dalton} program, we employ the default settings, but explore two different guesses for the starting orbitals, namely, the default guess in \textsc{Dalton} and the natural orbitals obtained with second-order M\o ller-Plesset perturbation theory (MP2), as suggested by the \textsc{Dalton} user manual.\cite{Dalton} The state-averaged and linear response calculations are performed with the CFOUR program. All the calculations in this work are done without enforcing point-group symmetry.
	
	\medskip
	
	\par\noindent\textbf{Analysis techniques for state-specific results.} To investigate the nature of state-specific solutions, we come up with two strategies. 
	
	\par\noindent (a) \textit{Singular Value Decomposition (SVD) analysis.} For any state to be analyzed, we compute the SVD of the difference between its 1-RDM and the ground-state 1-RDM, both expressed in the MO basis of the ground state. Looking at the obtained singular values offers a way to understand the excitation level of the considered state. In particular, if two singular values are found to be close to one while the others are close to zero, the occupations of two MOs differ by approximately one in the considered saddle point with respect to the occupations of the ground state. The analyzed state can therefore be seen as a good candidate representing a single excitation with respect to the ground state. Similarly, if two singular values are close to two while the others remain near zero, the state likely corresponds to a double excitation with respect to the ground state. One has to be aware of the fact that the non-zero singular values might not be exactly one or two because the two states in comparison do not necessarily share the same set of MOs. 
	
	\par
\begin{figure*}[!t]
	\centering
	\begin{subfigure}{0.48\linewidth}
		\centering
		\includegraphics[width=\columnwidth]{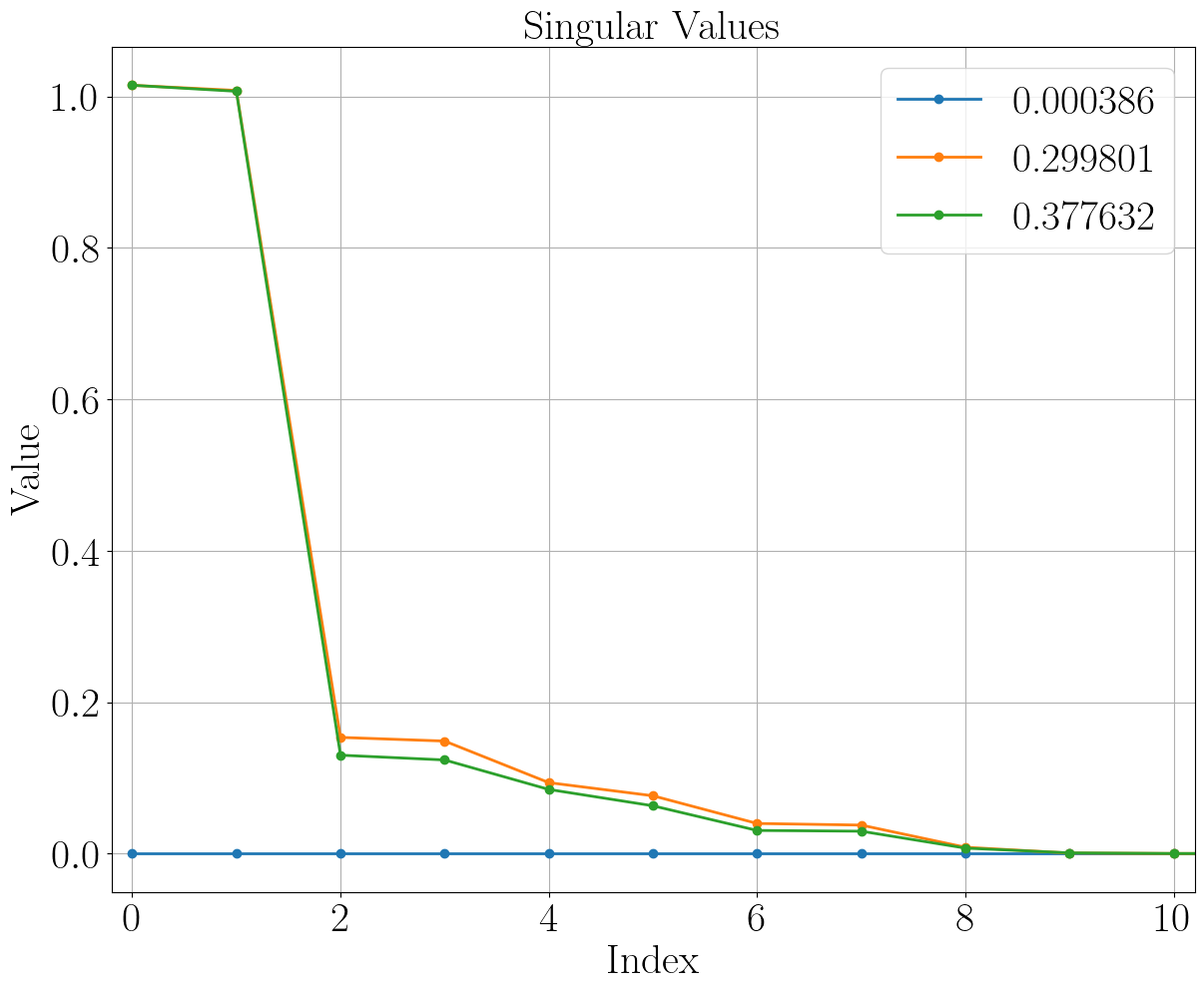}
		\caption{Water with CAS(8,6)/6-31G.}
		\label{fig:water_phys}
	\end{subfigure}
	\begin{subfigure}{0.48\linewidth}
		\centering
		\includegraphics[width=\columnwidth]{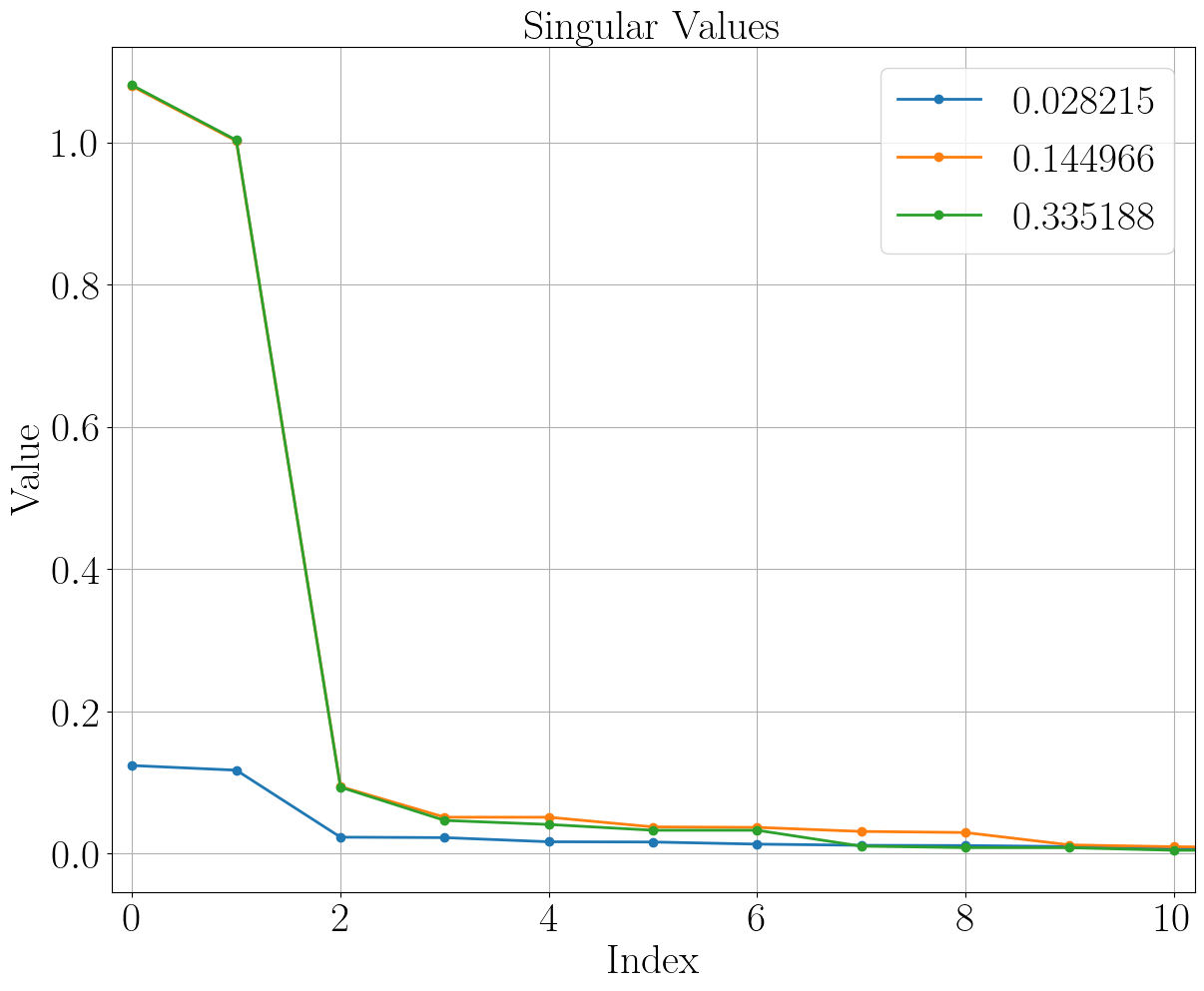}
		\caption{Formaldehyde with CAS(4,3)/6-31G.}
		\label{fig:form_phys}
	\end{subfigure}
	
	\medskip
	\begin{subfigure}{0.48\linewidth}
		\centering
		\includegraphics[width=\columnwidth]{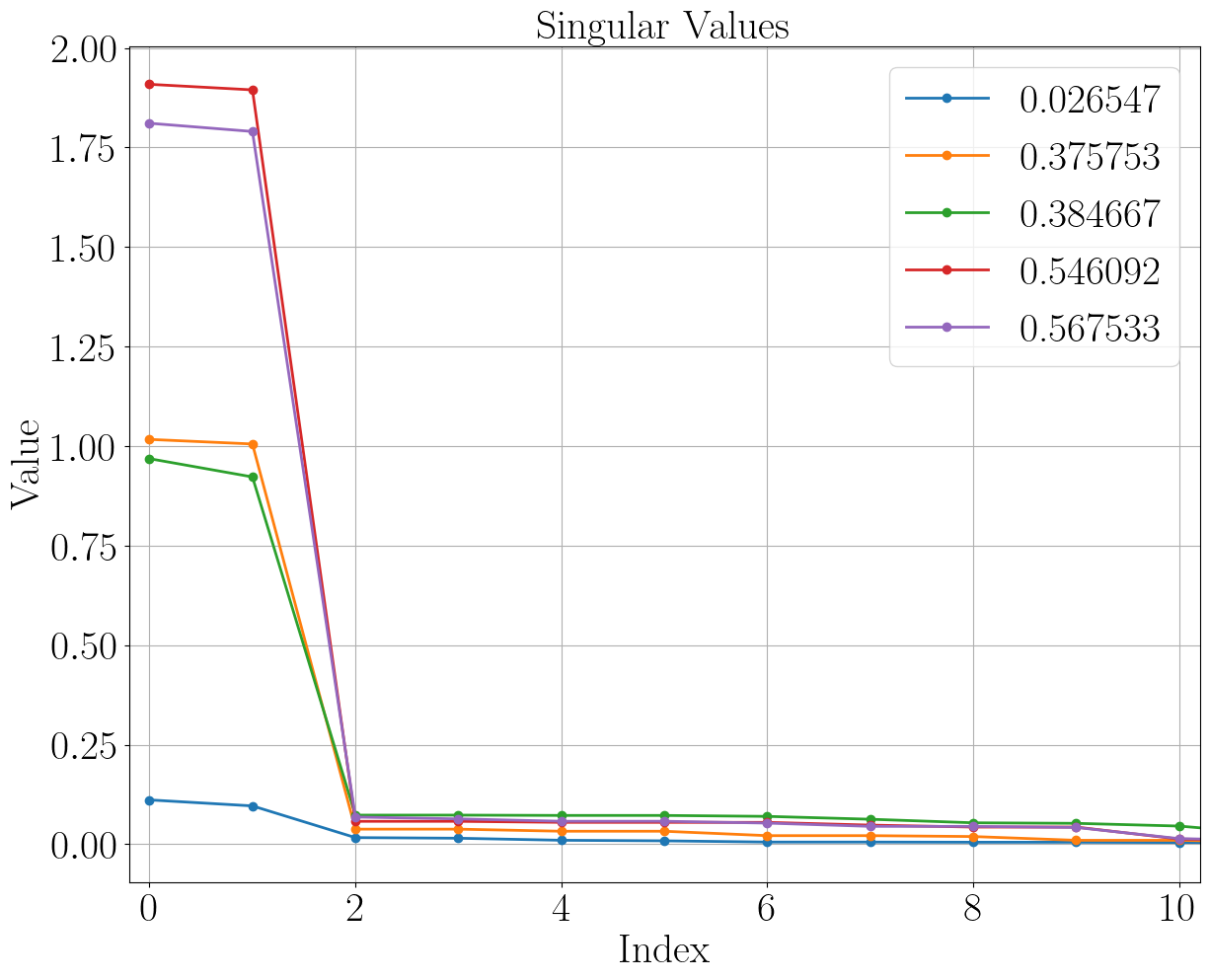}
		\caption{Ethylene with CAS(2,2)/6-31G.}
		\label{fig:ethylene_phys}
	\end{subfigure}
	\caption{SVD analysis for some of state-specific solutions for the three molecules. Singular values are associated with the differences between the 1-RDMs of the states with the energy differences shown in the legends and the 1-RDM of the ground state. Both 1-RDMs are expressed in the MO basis of the ground state.}
	\label{fig:svd}
\end{figure*}

\noindent (b) \textit{Eigenvector analysis.} Due to the nonlinearity of the CASSCF parametrization, it is highly possible that the saddle point characterized by singular values of one or two from the SVD analysis is not unique (see Fig.~\ref{fig:ethylene_phys} below for an example). In order to find the best approximation for the excited states, we analyze the eigenvectors of the Riemannian Hessian associated with negative eigenvalues, which can be readily obtained from the description of the CGAM (cf. Section \ref{sec:CASSCF SS}). More precisely, we calculate the norms of the $d$ and $\kappa$ components of these eigenvectors, as these describe the magnitudes of the rotations in the configuration and orbital spaces, respectively. When using CASSCF theory, if an excitation is characterized by a different occupation of the orbitals in the CAS space (which could be the case for lowest excitations), the excited state should be characterized by $\Vert d\Vert$'s significantly larger than $\Vert\kappa\Vert$'s, showing that its Morse index is mainly carried by the CI part. On the other hand, spurious solutions are characterized by smaller $\Vert d\Vert$'s relative to $\Vert\kappa\Vert$'s, showing that they arise from the rotation of the orbitals. However, if the CAS space is insufficient (especially for higher excitations), the analysis would become more subtle. In this case, $\Vert d\Vert$ and $\Vert\kappa\Vert$ could even be of comparable magnitude, and we pick out the solution with the largest $\Vert d\Vert$.

	\medskip
	
	\begin{table*}[!t]
		\centering
		\caption{State-specific results on the molecules examined in this paper. The ground-state energies are respectively -76.039557 $E_h$ for water, -113.849431 $E_h$ for formaldehyde, and -78.034921 $E_h$ for ethylene. The ``$\Delta E$'' shows the energy difference from the ground state. For the ``\textsc{Dalton}-NEO'' column, the ``Morse ind.'' lists the identified Morse index of the state and the ``HF, MP2'' indicates whether the initialization is based on HF and MP2 calculations or not. For the ``CFOUR-CGAM'' column, the ``Initial'' shows which CASCI root is taken for initialization.}
		\label{tab:state-specific}
		\begin{subtable}{\linewidth}
			\centering
			\caption{Water molecule with CAS(8,6)/6-31G.}
			\begin{threeparttable}
				\begin{tabular}{c||ccc||ccc}
					\hline\hline
					\multirow{2}{*}{\tabincell{c}{Target\\ex. state}} & \multicolumn{3}{c||}{\textsc{Dalton}-NEO} & \multicolumn{3}{c}{CFOUR-CGAM} \\\cline{2-7}
					& \multicolumn{1}{c}{Initial} & \multicolumn{1}{c}{$\Delta E$ ($E_h$)} & \multicolumn{1}{c||}{Morse ind.} & \multicolumn{1}{c}{Initial} & \multicolumn{1}{c}{$\Delta E$ ($E_h$)} & Morse ind. \\\hline
					1 & MCSCF & 0.299801\tnote{+} & 1 & CASCI-gs & 0.000386 & 1\\
					& MCSCF+HF+MP2 & 0.377632\tnote{$\dagger$} & 2 & CASCI-1ex & 0.299801\tnote{+} & 1\\\hline
					2 & MCSCF+HF+MP2 & 0.301031 & 2 & CASCI-2ex & 0.377632\tnote{$\dagger$} & 2\\
					& MCSCF & 0.401862 & 2 &  &  & \\
					\hline\hline
				\end{tabular}
			\begin{tablenotes}
				\footnotesize
				\item[+] HOMO-LUMO single excitation.
				\item[$\dagger$] HOMO-LUMO+1 single excitation.
			\end{tablenotes}
		\end{threeparttable}
	\end{subtable}
	
	\medskip
	
	\begin{subtable}{\linewidth}
		\centering
		\caption{Formaldehyde molecule with CAS(4,3)/6-31G.}
		\begin{threeparttable}
				\begin{tabular}{c||ccc||ccc}
					\hline\hline
					\multirow{2}{*}{\tabincell{c}{Target\\ex. state}} & \multicolumn{3}{c||}{\textsc{Dalton}-NEO} & \multicolumn{3}{c}{CFOUR-CGAM} \\\cline{2-7}
					& Initial & \multicolumn{1}{c}{$\Delta E$ ($E_h$)} & \multicolumn{1}{c||}{Morse ind.} & \multicolumn{1}{c}{Initial} & \multicolumn{1}{c}{$\Delta E$ ($E_h$)} & Morse ind. \\\hline
					1 & MCSCF & 0.146472 & 2 & CASCI-gs & 0.028215 & 1\\
					& MCSCF+HF+MP2 & 0.338464 & 3 & CASCI-1ex & 0.144966\tnote{+} & 1\\\hline
					2 & MCSCF & 0.338464 & 3 & CASCI-2ex & 0.146478 & 2 \\
					& MCSCF+HF+MP2 & 0.457140 & 7 & random & 0.335188\tnote{$\dagger$} & 2\\
					\hline\hline
				\end{tabular}
			\begin{tablenotes}
				\footnotesize
				\item[+] $n\to\pi^*$ single excitation.
				\item[$\dagger$] $\pi\to\pi^*$ single excitation.
			\end{tablenotes}
		\end{threeparttable}
	\end{subtable}
	
	\medskip
	
	\begin{subtable}{\linewidth}
		\centering
		\caption{Ethylene molecule with CAS(2,2)/6-31G.}
		\begin{threeparttable}
				\begin{tabular}{c||ccc||ccc}
					\hline\hline
					\multirow{2}{*}{\tabincell{c}{Target\\ex. state}} & \multicolumn{3}{c||}{\textsc{Dalton}-NEO} & \multicolumn{3}{c}{CFOUR-CGAM} \\\cline{2-7}
					& Initial & \multicolumn{1}{c}{$\Delta E$ ($E_h$)} & \multicolumn{1}{c||}{Morse ind.} & \multicolumn{1}{c}{Initial} & \multicolumn{1}{c}{$\Delta E$ ($E_h$)} & Morse ind. \\\hline
					1 & MCSCF & 0.379889 & 2 & CASCI-gs & 0.026547 & 1 \\
					& MCSCF+HF+MP2 & 0.394747 & 1 & CASCI-1ex & 0.026571 & 1 \\
					&  &  &  & random & 0.384667\tnote{+} & 1\\\hline
					2 & MCSCF+HF+MP2 & 0.571481 & 3 & CASCI-2ex & 0.027343 & 2\\
					& MCSCF & 0.796880 & 3 & random & 0.567533\tnote{++} & 2\\
					\hline\hline
				\end{tabular}
			
			\begin{tablenotes}
				\footnotesize
				\item[+] HOMO-LUMO single excitation.
				\item[++] HOMO-LUMO double excitation.
			\end{tablenotes}
		\end{threeparttable}
	\end{subtable}
\end{table*}

\par\noindent\textbf{Results and discussions.} We report in Table~\ref{tab:state-specific} the state-specific results on the three molecules. The table shows the comparison between the \textsc{Dalton}-NEO and CFOUR-CGAM results, with different initialization strategies. For the former one, we also report the identified Morse indices of the obtained states. The index is identified by 
performing a CASCI calculation with the orbitals converged by \textsc{Dalton} and locating the root corresponding to the NEO solution, finally calculating the eigenvalues of the Riemannian Hessian and counting the number of negative ones. Note that we augment the CFOUR-CGAM results with those from random searches only when the chemically motivated initializations do not lead to physically meaningful states. By using the analysis techniques detailed above, we attributed the physical nature of the obtained state-specific solutions. Our analysis is detailed below.


\begin{figure*}[!t]
	\centering
	\begin{subfigure}{0.48\linewidth}
		\centering
		\includegraphics[width=\columnwidth]{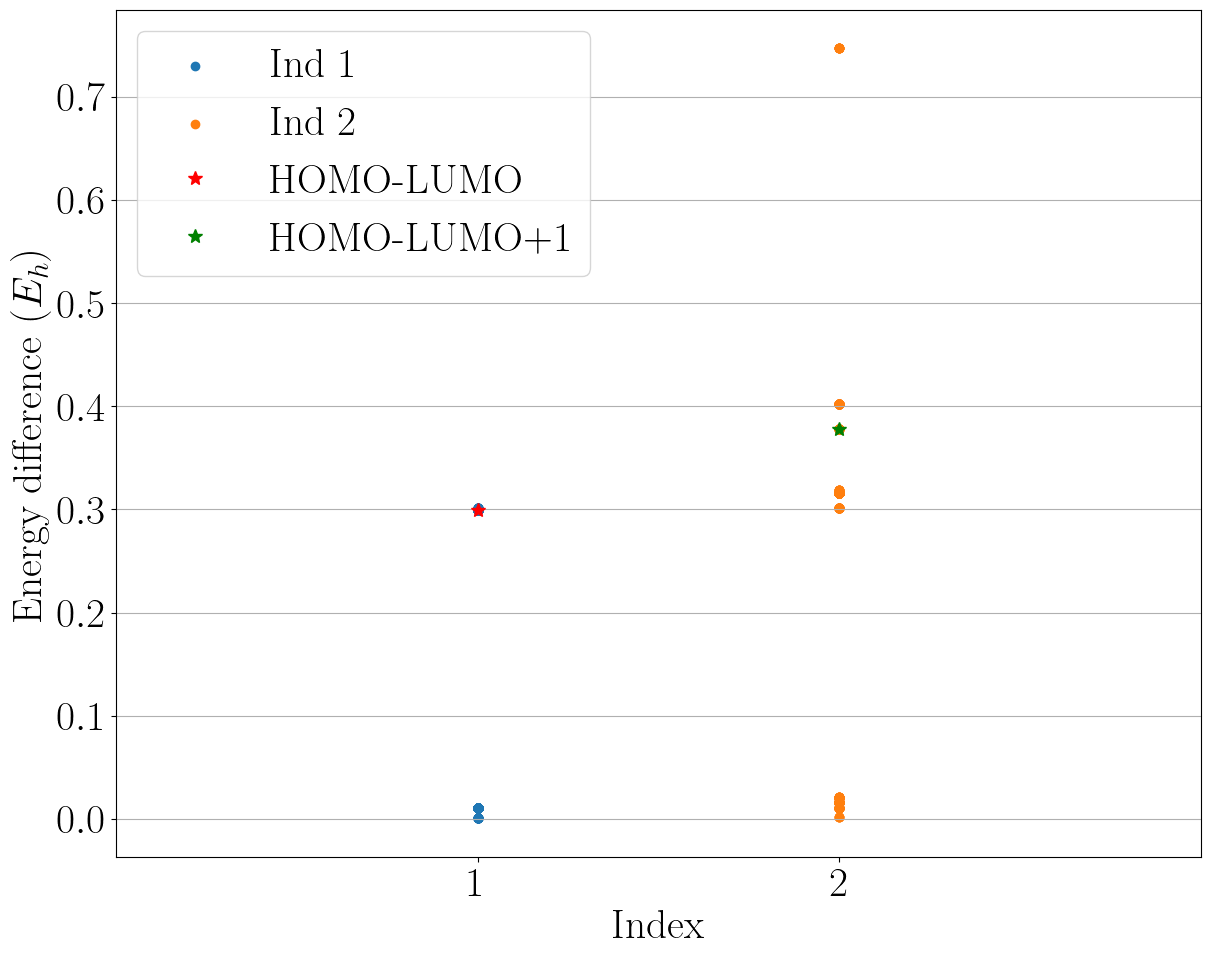}
		\caption{Water with CAS(8,6)/6-31G.}
		\label{fig:water_random}
	\end{subfigure}
	\begin{subfigure}{0.48\linewidth}
		\centering
		\includegraphics[width=\columnwidth]{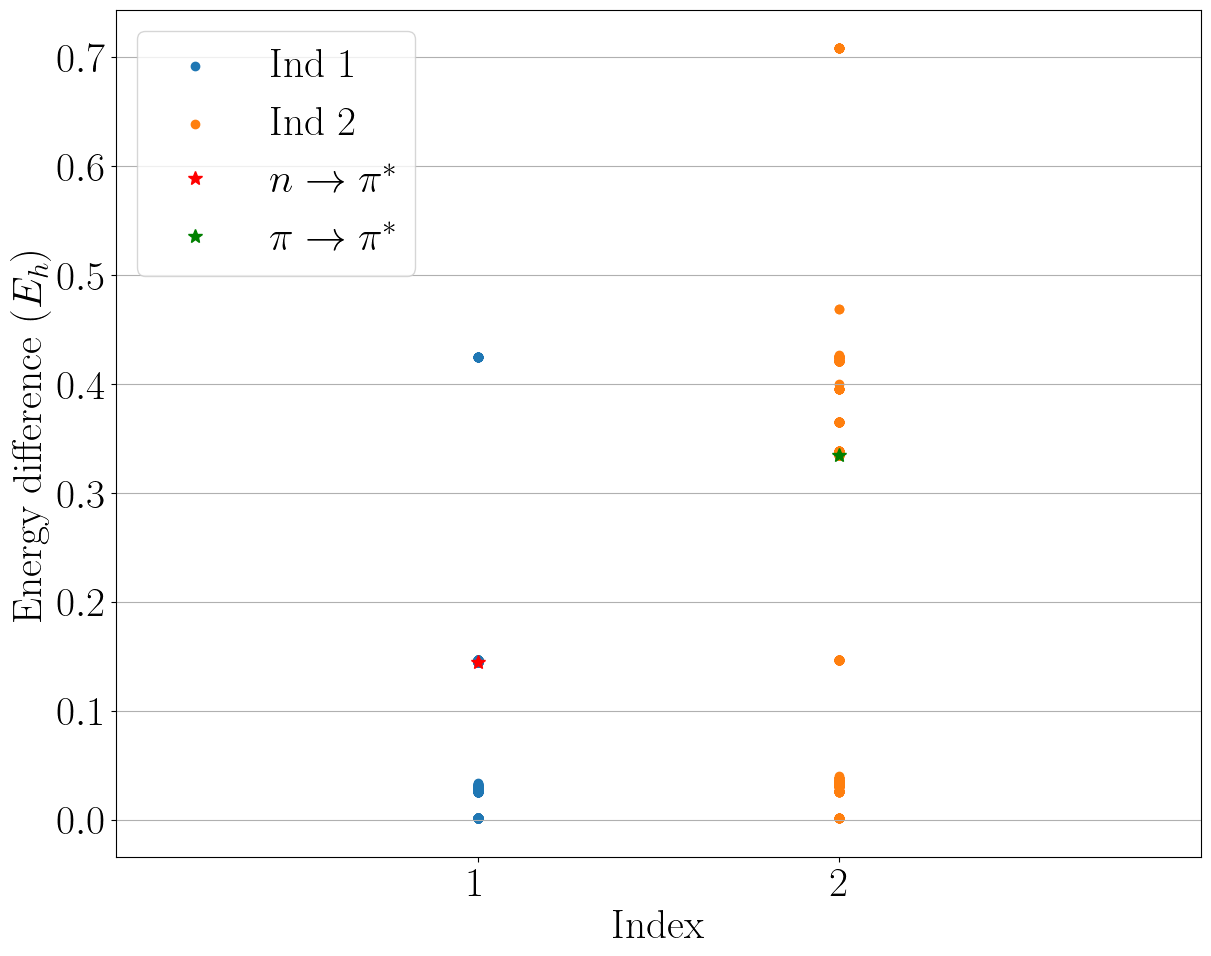}
		\caption{Formaldehyde with CAS(4,3)/6-31G.}
		\label{fig:form_random}
	\end{subfigure}
	
	\medskip
	\begin{subfigure}{0.48\linewidth}
		\centering
		\includegraphics[width=\columnwidth]{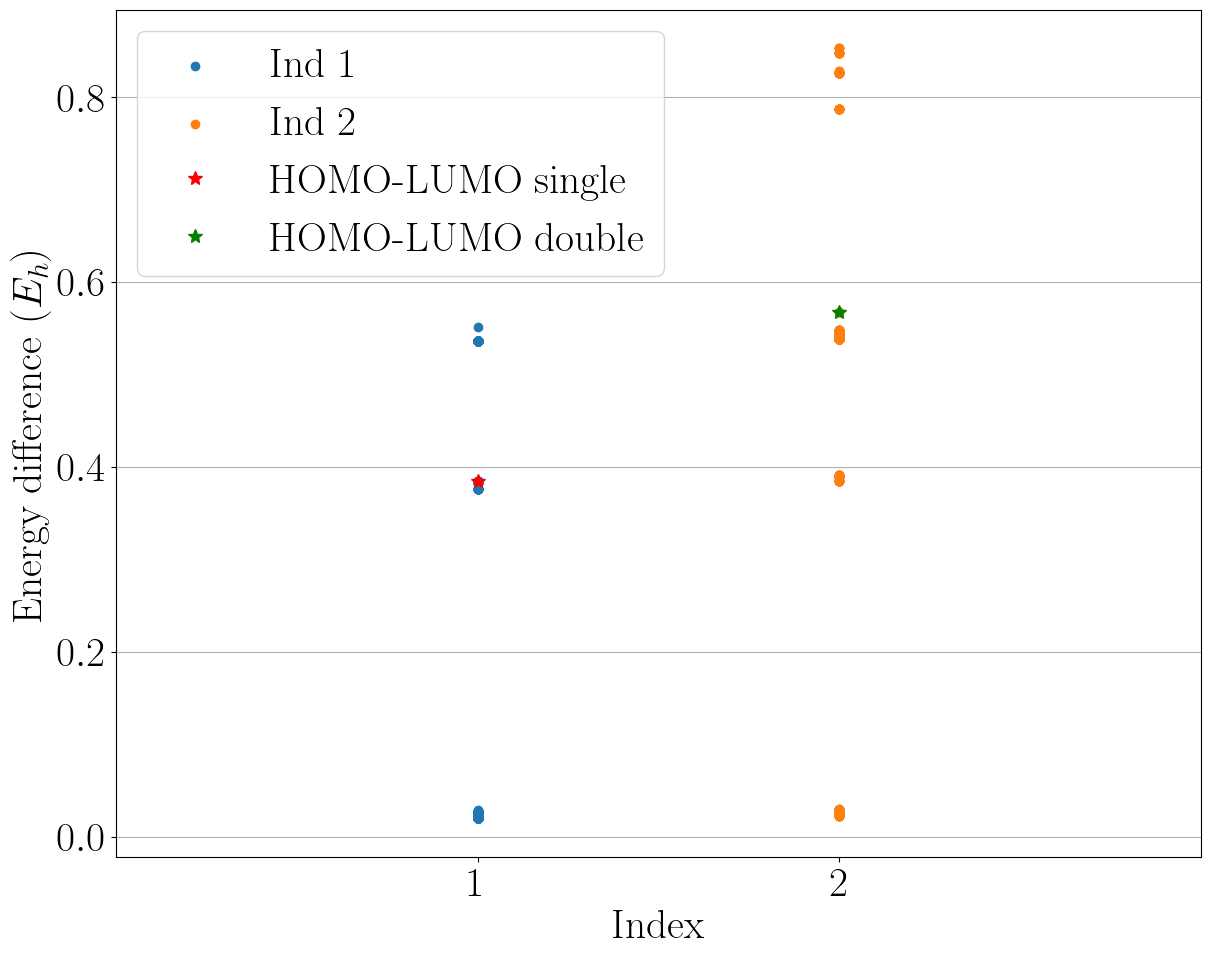}
		\caption{Ethylene with CAS(2,2)/6-31G.}
		\label{fig:ethylene_random}
	\end{subfigure}
	\caption{Energy differences of the index-1 and index-2 saddle points for the analyzed molecules found through 500 random searches with the developed CGAM. The stars indicate which cluster of saddle points corresponds to the physical excitation.}
	\label{fig:random}
\end{figure*}

\par Fig.~\ref{fig:svd} shows the results from the SVD analysis for some of the state-specific solutions. The solutions with small energy differences from the ground state are indeed spurious solutions, characterized by all singular values close to zero. The states that we identified as single excitations in Table \ref{tab:state-specific} are indeed characterized by two singular values close to one: the HOMO-LUMO single excitations for the water, formaldehyde, and ethylene molecules with respective excitation energies 0.299801 $E_h$, 0.144966 $E_h$, and 0.384667 $E_h$ are described by index-1 saddle points; the HOMO-LUMO+1 single excitation for the water molecule with the excitation energy 0.377632 $E_h$ and the $\pi\to\pi^*$ single excitation for the formaldehyde molecule with the excitation energy 0.335188 $E_h$ are described by index-2 saddle points. The HOMO-LUMO double excitation of ethylene with the excitation energy 0.567533 $E_h$, reported in Table~\ref{tab:state-specific}, corresponds to an index-2 saddle point and is characterized by two singular values close to two. 

\par We point out that the SVD analysis does not suffice to analyze the energy landscape in general, but could be complemented by the eigenvector analysis. One typical example where the SVD does not fully characterize the state is shown in Fig.~\ref{fig:ethylene_phys}, where we observe that there are two good candidates for both single and double excitations. For the single excitation, the eigenvector analysis shows that the state at 0.384667 $E_h$ has $\Vert d\Vert\gg\Vert\kappa\Vert$, therefore identifying the HOMO-LUMO transition; on the other hand, the state at 0.375753 $E_h$ has $\Vert d\Vert\ll\Vert\kappa\Vert$, identifying a spurious saddle point. For the double excitation of ethylene, the eigenvector analysis shows that $\Vert d\Vert$'s and $\Vert\kappa\Vert$'s are of comparable magnitude in two eigenvectors for the state with the energy difference 0.567533 $E_h$. The state with the energy difference 0.546092 $E_h$, which is also with singular values close to two, is characterized by $\Vert d\Vert$'s much smaller than $\Vert\kappa\Vert$'s. The eigenvector analysis thus leads us to identify the state at 0.567533 $E_h$ as the second excited states. 

\par From Table \ref{tab:state-specific} and Fig. \ref{fig:svd}, we observe that the chemically motivated initial guesses do not always lead to physical excited states. 
In many cases, the CASCI states do not appear to be in the attractive basin of any excited state and guide the CGAM to the nearest spurious solution. These spurious solutions are inevitably present due to the inherent nonlinearity introduced by the CASSCF theory. When the designated initializations are not reliable, the robustness of the developed CGAM could be made use of by adopting the random initialization strategy, to find saddle points with a prefixed target Morse index on the CASSCF energy landscapes. This methodology allowed us to locate the excited states describing the $\pi\rightarrow\pi^*$ single excitation of formaldehyde and the HOMO-LUMO single and double excitations of ethylene. Nevertheless, these successes do not come at no price: due to the nonlinearity of the CASSCF theory, the CASSCF energy landscape could have plenty of spurious saddle points with the target Morse index, which renders the random search process lengthy and necessitates careful quantum chemistry analysis for each obtained solution. 

\par This difficulty can be visualized in Fig.~\ref{fig:random}. For each of the inspected molecules, we performed 500 calculations with different random initializations. The energy differences of the found saddle points from the ground state are here shown. For water (Fig.~\ref{fig:water_random}), our calculations showed the presence of mainly three clusters of index-1 and seven clusters of index-2 saddle points; formaldehyde (Fig.~\ref{fig:form_random}) exhibits four clusters of index-1 and nine clusters of index-2 saddle points; the analysis on ethylene (Fig.~\ref{fig:ethylene_random}) leads to five clusters of index-1 and seven clusters of index-2 saddle points. Some index-2 saddle points are found energetically close to index-1 ones, which can be understood by considering that index-2 saddle points can be spurious critical points approximating the first excited state. In our analysis, we were able to identify the physical excitations by combining the SVD and the eigenvector arguments detailed above.

\par Comparing the CGAM results with the NEO ones, we notice that the NEO algorithm often converges to different saddle points, and the identified Morse indices of the given solutions do not always correspond to the targeted ones. The NEO algorithm is a second-order optimization method based on a trust-region, level-shifted Newton strategy, and is in general known for its robustness and stability.\cite{jensen1986direct,jensen1984direct} Nevertheless, not only the state to which it converges strongly depends, as one can expect, on the starting point, as can be seen by comparing the results obtained with and without the MP2 natural orbitals as a guess, but the saddle point obtained can easily exhibit a Morse index different from the one sought. This does not mean that \textsc{Dalton}'s results are in any way spurious - indeed, we have never observed solutions that are very close to the ground state - but that the straightforward association between the Morse index of the saddle point and the order of the excited state, which holds exactly for the exact theory, should be taken with a grain of salt. 
\par Starting from water, we can see that NEO and CGAM both identify the HOMO-LUMO and HOMO-LUMO+1 excitations, but both experience some difficulties. NEO finds both by seeking index-1 saddle points employing different guesses, contrary to CGAM that correctly finds the former while looking for index-1 saddle points, and the latter for index-2 ones. On the other hand, CGAM finds a spurious solution, very close to the ground state, as an index-1 saddle point. When looking for index-2 saddle points, NEO finds two solutions that we could not associate to the desired states. In particular, the first one, at 0.301031 $E_h$, has Morse index two, but is very close to the first excited states and is likely a spurious solution. 
\par Things are even more complicated for formaldehyde. Here, it is interesting to notice that both NEO and CGAM find very close energies for the $n\rightarrow\pi^*$ and $\pi\rightarrow\pi^*$ states, however, NEO identifies them as index-2 and index-3 saddle points, while CGAM as index-1 and index-2 ones, respectively. This does not mean that CGAM's solutions are better or more accurate, though, as it is quite likely that, as we work without enforcing point-group symmetry, CGAM has found a lower, symmetry broken solution, i.e., CGAM's solutions could be connected to NEO's ones by a symmetry breaking instability. While we are working without enforcing symmetry, it is in fact likely that the correct symmetry is numerically enforced in NEO. While this could be the case for our calculations when starting from a deterministic guess, it is clearly not the case for the random initialization that was required to find the second excited state. Again, we note that NEO finds the second excited state both by looking for the second excitation using the default guess and by looking for the first one starting from MP2 natural orbitals, which further points out how the results delicately depend on the initial guess. 
\par The results for ethylene further demonstrate that using the Morse index as an indication of the excitation level can be misleading. Seeking an index-1 saddle point, NEO finds two different solutions starting with or without MP2 natural orbitals as a guess, but interestingly, the index-2 saddle point has an energy that is lower than the index-1 one, which clearly shows how dangerous it is to associate the saddle point Morse indices with excitation orders. Again, we note that CGAM and NEO managed to find somewhat similar excitation energies (0.384667 vs 0.379889 or 0.394747$E_h$ for the first excited state and 0.567533 vs 0.571481$E_h$ for the second one, respectively). Again, we note the different Morse indices of the NEO solutions, which could be due to a symmetry breaking instability. This is particularly likely as using a deterministic guess leads only to spurious solutions for CGAM, which forced us to employ a 
random initialization strategy, which is very likely to produce symmetry broken solutions. 



\begin{table}[!t]
	\centering
	\caption{Excitation energies $\Delta E$ (in $E_h$) obtained from State-Specific (SS), State-Averaged (SA), and linear response (LR) calculations for the water, formaldehyde and ethylene molecules. For each molecule, the two lowest-lying excited states are listed. SA calculations were carried out averaging over 3 states for all systems.}
	\label{tab:excitation energies SS SA LR}
	\begin{tabular}{c||ccc}
		\hline\hline
		No. & SS & SA & LR \\
		\hline
		
		\multicolumn{4}{c}{Water with CAS(8,6)/6-31G}\\
		\hline
		1  & ~~0.299801~~ &  ~~0.279149~~ & ~~0.313274~~ \\
		2  & ~~0.377632~~ &  ~~0.357262~~ & ~~0.388773~~ \\
		\hline
		\multicolumn{4}{c}{Formaldehyde with CAS(4,3)/6-31G}\\
		\hline
		1 & ~~0.144960~~ &  ~~0.147317~~ & ~~0.138848~~ \\
		2 & ~~0.335188~~ &  ~~0.464643~~ & ~~0.325269~~ \\
		\hline
		\multicolumn{4}{c}{Ethylene with CAS(2,2)/6-31G}\\
		\hline
		1  & ~~0.384667~~ &  ~~0.388940~~ & ~~0.354318~~ \\
		2  & ~~0.567533~~ &  ~~0.568822~~ & ~~0.562656~~ \\
		
		\hline\hline
	\end{tabular}
\end{table}

In Table~\ref{tab:excitation energies SS SA LR}, we compare the excitation energies of the first two excited states obtained from state-specific (SS) calculations with those obtained in the state-averaged (SA) and linear response (LR) framework. We observe that the SS results are always in the range between SA and LR, therefore testifying the qualitative correctness of our calculations. While for the water molecule, the LR excitation energies are the highest, for formaldehyde and ethylene the LR results are the lowest. For all inspected excitations, our results agree with those found with the most widely used approximations in the CASSCF framework. We also point out that the characterization of the doubly excited state of ethylene, as shown in Fig.~\ref{fig:ethylene_phys}, is supported by the comparison with the SA and LR excitation energies.

In conclusion, the numerical experiments performed in this section serve as a cautionary tale for state-specific CASSCF excited state calculations. Genuine excited states can appear as saddle points with a Morse index that does not correspond to the excitation level, and the Morse indices and energy orders can even be different. Spurious states can be easily encountered and brute-force approaches, such as using a large number of random guesses, may be required in order to find the desired solution. In other words, state-specific CASSCF is in no way a black-box tool to locate and characterize excited states.
Nevertheless, the developed CGAM provides a powerful tool to target specific eigenstates with a given Morse index, therefore hoping to approximate the excited states. Our SVD analysis showed that the energy landscapes even of simple molecules are quite complex, leading to a large number of saddle points per Morse index. The identification of spurious states is therefore fundamental to obtain accurate physical results. We shall also remark that the quality of CASSCF results depends on many factors, one of them being the choice of the CAS space, and the reported excitation energies can only be interpreted qualitatively, more than quantitatively.

\section{Conclusions}

\par In this paper, we propose a unified K\"ahler manifold-based framework for describing electronic excitation energies within the CASSCF theory, encompassing both state-specific and linear response approaches. Different from HF, DFT, and FCI, which are addressed in Part I of this series of contributions,\cite{grazioli2026critical} CASSCF exhibits a much more complicated manifold structure, arising from the coupling between CI and orbital degrees of freedom. By analyzing the K\"ahler structure of the CASSCF manifold, we derive the CASSCF Hamiltonian dynamics, and provide straightforward and systematic formulations of CASSCF linear response equations\cite{olsen1985linear} and stationary conditions, the latter being central to state-specific methods. Equipped with the geometrical tools and taking the energy landscape point of view, this paper also presents a robust yet efficient state-specific method, named CGAM. Compared with existing ones, the CGAM does not require expensive second-order information and is able to converge to saddle points of prefixed Morse indices from any initial guess. 


\par We ran the developed algorithm on water with CAS(8,6), formaldehyde with CAS(4,3), and ethylene with CAS(2,2) to find index-1 and index-2 saddle points of the corresponding energy landscapes. We notice that chemically reasonable initializations (based on CASCI or MP2 calculations) can be good guesses in some cases, but are not always reliable. This is shown in the case of ethylene, where only random searches allowed us to locate physically meaningful excited states. This example also demonstrates the difficulty in navigating the CASSCF energy landscape even for simple examples, on which there lie plenty of spurious critical points (i.e., those not corresponding to physical states), due to the nonlinearity introduced by the CASSCF theory. The identification of excited states among them was possible only after careful post-processing, e.g., SVD analysis and eigenvector analysis conducted in this work.

\par Building on our algorithmic developments and on the mentioned analysis tools, we successfully located meaningful excited states for the three molecules, such as HOMO-LUMO single/double excitations and $n\to\pi^*$, $\pi\to\pi^*$ single excitations. 
The qualitative correctness of the state-specific results is validated by further comparing them with those from state-averaged and linear response calculations. 


\par Concluding this study, we highlight the complexity of using state-specific calculations for the CASSCF method, even when studying small molecules with small basis sets. At the current stage, any state-specific algorithm cannot be used as a black-box tool, as among the saddle points, which are all mathematically admissible solutions, the physically correct excited states can be identified only after careful analysis, involving tools from both differential geometry and quantum chemistry. 
This situation could be improved by designing methods that navigate the CASSCF energy landscape more efficiently (see, e.g., Ref. \citenum{lelievre2024using} for the case of a flat space), by developing automatic yet reliable tools for analyzing states, and by turning Lewin's theoretical global characterizations of physically-admissible multiconfiguration excited states\cite{lewin2004} into practical algorithms (see Refs.~\citenum{cances2006,lewin2008} for an application to the simple case of the first excited state of the H$_2$ molecule). These research directions will be explored in future works.


\section*{Acknowledgements}

This project has received funding from the European Research Council (ERC) under the European Union’s Horizon 2020 research and innovation program (grant agreement EMC2 No 810367). This work has benefited from French State support managed by ANR under the France 2030 program through the MaQui CNRS Risky and High-Impact Research program (RI)$^2$ (grant agreement ANR-24-RRII-0001). 
T.N. and F.L. further acknowledge support from ICSC-Centro Nazionale di Ricerca in High Performance Computing, Big Data, and Quantum Computing, funded by the European Union-NextGenerationEU-PNRR, Missione 4 Componente 2 Investimento 1.4.

\section{Conflicts of interest}
The authors have no conflict to disclose.

\section{Author declarations}
\subsection{Author contributions}
Laura Grazioli: conceptualization, investigation, methodology, software, validation, writing\\
Yukuan Hu: conceptualization, investigation, methodology, software, validation, writing\\
Tommaso Nottoli: conceptualization, investigation, methodology, software, validation, writing\\
Filippo Lipparini: conceptualization, investigation, methodology, funding acquisition, project administration, resources, supervision, writing.\\
Eric Cancès: conceptualization, investigation, methodology, funding acquisition, project administration, resources, supervision, writing.

\subsection{Data availability}
The data that support the findings of this study are available within the article.

\section{ORCID}

Laura Grazioli https://orcid.org/0000-0002-9404-8587

Yukuan Hu https://orcid.org/0000-0002-5372-3814

Tommaso Nottoli https://orcid.org/0000-0002-9543-6127

Filippo Lipparini https://orcid.org/0000-0002-4947-3912

Eric Cancès https://orcid.org/0000-0002-8876-8254

\appendix

\section{Hermitian inner product on the CASSCF manifold}\label{appsec:CASSCF manifold inner product}

We derive the Hermitian inner product in Eq. \eqref{eqn:CASSCF manifold inner product}. First note that for $[(c,C)]\in\mathcal M$ and $(d,\kappa)\in T_{[(c,C)]}\mathcal M$, 
$$d_{[(c,C)]}\pi(d,\kappa)=\bigg(\sum_{A=1}^{N_{\det}}d_A\ket{A^C}+\sum_{p>q}\kappa_{pq}\widehat p^\dagger\widehat q\ket{0}\bigg)\bra{0}+{\rm h.c.}$$
Moreover, the states in different groups $\{\ket{A^C}\}_{A=1}^{N_{\det}}$, $\{\widehat u^\dagger\widehat i\ket{0}\}_{(u,i)\in{\rm AI}}$, $\{\widehat a^\dagger\widehat i\ket{0}\}_{(a,i)\in{\rm EI}}$, and $\{\widehat a^\dagger\widehat u\ket{0}\}_{(a,u)\in{\rm EA}}$ are mutually orthogonal to each other. Therefore, for any $(d,\kappa),(d',\kappa')\in T_{[(c,C)]}\mathcal M$, 
\begin{align*}
	&\braket{d_{[(c,C)]}\pi(d,\kappa)|d_{[(c,C)]}\pi(d',\kappa')}=\\
	&=2\bigg(\sum_{A,A'=1}^{N_{\det}}d_A^*d_{A'}'\braket{A^C|A^{'C}}+\sum_{u,v}\sum_{i,j}\kappa_{ui}^*\kappa_{vj}'\braket{0|\widehat i^\dagger\widehat u\widehat v^\dagger\widehat j|0}\\
	&+\sum_{a,b}\sum_{i,j}\kappa_{ai}^*\kappa_{bj}'\braket{0|\widehat i^\dagger\widehat a\widehat b^\dagger\widehat j|0}+\sum_{a,b}\sum_{u,v}\kappa_{au}^*\kappa_{bv}'\braket{0|\widehat u^\dagger\widehat a\widehat b^\dagger\widehat v|0}\bigg).
\end{align*}
The first term on the right-hand side is $d^\dagger d'$. For the second term, by the properties of creation and annihilation operators in second quantization, 
$$\braket{0|\widehat i^\dagger\widehat u\widehat v^\dagger\widehat j|0}=\delta_{ij}\braket{0|\widehat u\widehat v^\dagger|0}=\delta_{ij}(\delta_{uv}-[\gamma_{c,C}^{\rm act}]_{vu}),$$
we obtain
\begin{align*}
	\sum_{u,v}\sum_{i,j}\kappa_{ui}^*\kappa_{vj}'\braket{0|\widehat i^\dagger\widehat u\widehat v^\dagger\widehat j|0}&=\sum_{i}\sum_{u,v}\kappa_{ui}^*(\delta_{uv}-[\gamma_{c,C}^{\rm act}]_{vu})\kappa_{vi}'\\
	&={\rm Tr}\big((\kappa_{\rm AI}')^\top(I-\gamma_{c,C}^{\rm act})\kappa_{\rm AI}^*\big).
\end{align*}
Similarly, we have for the third and last terms that
\begin{align*}
	\sum_{a,b}\sum_{i,j}\kappa_{ai}^*\kappa_{bj}'\braket{0|\widehat i^\dagger\widehat a\widehat b^\dagger\widehat j|0}&={\rm Tr}\big((\kappa_{\rm EI}')^\top\kappa_{\rm EI}^*\big),\\
	\sum_{a,b}\sum_{u,v}\kappa_{au}^*\kappa_{bv}'\braket{0|\widehat u^\dagger\widehat a\widehat b^\dagger\widehat v|0}&={\rm Tr}\big(\kappa_{\rm EA}^*\gamma_{c,C}^{\rm act}(\kappa_{\rm EA}')^\top\big).
\end{align*}
In all, we have the full expression in Eq. \eqref{eqn:CASSCF manifold inner product}.

\bibliographystyle{apsrev4-1}
\bibliography{ref}

\end{document}